\documentclass[twocolumn,letterpaper,aps,prc,superscriptaddress,nofootinbib,longbibliography,floatfix]{revtex4-1}

\usepackage{graphicx}	
\usepackage{multirow}
\usepackage{xspace}	

\newcommand{\pt}{\mbox{$p_T$}\xspace}

\newcommand{\Npart}{\mbox{$N_{\rm part}$}\xspace}

\newcommand{\meanpt}{\mbox{$\langle p_T \rangle$}\xspace}

\newcommand{\sqsn}{\mbox{$\sqrt{s_{_{NN}}}$}\xspace}
\newcommand{\psifvtxs}{\mbox{$\Psi_2^{\rm FVTXS}$}\xspace}
\newcommand{\psibbcs}{\mbox{$\Psi_2^{\rm BBCS}$}\xspace}
\newcommand{\psipp}{\mbox{$\Psi_2^{\rm Parton\ Plane}$}\xspace} 
\newcommand{\dndeta}{\mbox{$dN_{ch}/d\eta$}\xspace}

\newcommand{\vtep}{v_2\{\rm EP\}}
\newcommand{\vtpp}{v_2\{\rm Parton\ Plane\}}

\newcommand{\pp}{\mbox{$p$$+$$p$}\xspace}

\newcommand{\ppb}{\mbox{$p$$+$Pb}\xspace}
\newcommand{\dau}{\mbox{$d$$+$Au}\xspace}
\newcommand{\heau}{\mbox{$^3$He$+$Au}\xspace}
\newcommand{\pdheau}{\mbox{$p/d/^3$He$+$Au}\xspace}

\newcommand{\geant}{{\sc geant}-3\xspace}
\newcommand{\ampt}{\xspace{\sc ampt}\xspace}
\newcommand{\sonic}{{\sc sonic}\xspace}
\newcommand{\supersonic}{super{\sc sonic}\xspace}

\newcommand{\sigparton}{\mbox{$\sigma_{\rm parton}$}\xspace}

\begin{document}

\title{Measurements of azimuthal anisotropy and charged-particle 
multiplicity in $d$$+$Au collisions 
at $\sqrt{s_{_{NN}}}=200$, 62.4, 39, and 19.6 GeV}

\newcommand{\abilene}{Abilene Christian University, Abilene, Texas 79699, USA}
\newcommand{\augie}{Department of Physics, Augustana University, Sioux Falls, South Dakota 57197, USA}
\newcommand{\banaras}{Department of Physics, Banaras Hindu University, Varanasi 221005, India}
\newcommand{\barc}{Bhabha Atomic Research Centre, Bombay 400 085, India}
\newcommand{\baruch}{Baruch College, City University of New York, New York, New York, 10010 USA}
\newcommand{\bnlcoll}{Collider-Accelerator Department, Brookhaven National Laboratory, Upton, New York 11973-5000, USA}
\newcommand{\bnlphys}{Physics Department, Brookhaven National Laboratory, Upton, New York 11973-5000, USA}
\newcommand{\caucr}{University of California-Riverside, Riverside, California 92521, USA}
\newcommand{\charlesczech}{Charles University, Ovocn\'{y} trh 5, Praha 1, 116 36, Prague, Czech Republic}
\newcommand{\chonbuk}{Chonbuk National University, Jeonju, 561-756, Korea}
\newcommand{\cns}{Center for Nuclear Study, Graduate School of Science, University of Tokyo, 7-3-1 Hongo, Bunkyo, Tokyo 113-0033, Japan}
\newcommand{\colorado}{University of Colorado, Boulder, Colorado 80309, USA}
\newcommand{\columbia}{Columbia University, New York, New York 10027 and Nevis Laboratories, Irvington, New York 10533, USA}
\newcommand{\czechtech}{Czech Technical University, Zikova 4, 166 36 Prague 6, Czech Republic}
\newcommand{\debrecen}{Debrecen University, H-4010 Debrecen, Egyetem t{\'e}r 1, Hungary}
\newcommand{\elte}{ELTE, E{\"o}tv{\"o}s Lor{\'a}nd University, H-1117 Budapest, P{\'a}zm{\'a}ny P.~s.~1/A, Hungary}
\newcommand{\eszterhazy}{Eszterh\'azy K\'aroly University, K\'aroly R\'obert Campus, H-3200 Gy\"ongy\"os, M\'atrai \'ut 36, Hungary}
\newcommand{\ewha}{Ewha Womans University, Seoul 120-750, Korea}
\newcommand{\fsu}{Florida State University, Tallahassee, Florida 32306, USA}
\newcommand{\gsu}{Georgia State University, Atlanta, Georgia 30303, USA}
\newcommand{\hiroshima}{Hiroshima University, Kagamiyama, Higashi-Hiroshima 739-8526, Japan}
\newcommand{\howard}{Department of Physics and Astronomy, Howard University, Washington, DC 20059, USA}
\newcommand{\ihepprot}{IHEP Protvino, State Research Center of Russian Federation, Institute for High Energy Physics, Protvino, 142281, Russia}
\newcommand{\illuiuc}{University of Illinois at Urbana-Champaign, Urbana, Illinois 61801, USA}
\newcommand{\inrras}{Institute for Nuclear Research of the Russian Academy of Sciences, prospekt 60-letiya Oktyabrya 7a, Moscow 117312, Russia}
\newcommand{\instpasczech}{Institute of Physics, Academy of Sciences of the Czech Republic, Na Slovance 2, 182 21 Prague 8, Czech Republic}
\newcommand{\isu}{Iowa State University, Ames, Iowa 50011, USA}
\newcommand{\jaea}{Advanced Science Research Center, Japan Atomic Energy Agency, 2-4 Shirakata Shirane, Tokai-mura, Naka-gun, Ibaraki-ken 319-1195, Japan}
\newcommand{\jyvaskyla}{Helsinki Institute of Physics and University of Jyv{\"a}skyl{\"a}, P.O.Box 35, FI-40014 Jyv{\"a}skyl{\"a}, Finland}
\newcommand{\kek}{KEK, High Energy Accelerator Research Organization, Tsukuba, Ibaraki 305-0801, Japan}
\newcommand{\korea}{Korea University, Seoul, 136-701, Korea}
\newcommand{\kurchatov}{National Research Center ``Kurchatov Institute", Moscow, 123098 Russia}
\newcommand{\kyoto}{Kyoto University, Kyoto 606-8502, Japan}
\newcommand{\lawllnl}{Lawrence Livermore National Laboratory, Livermore, California 94550, USA}
\newcommand{\losalamos}{Los Alamos National Laboratory, Los Alamos, New Mexico 87545, USA}
\newcommand{\lund}{Department of Physics, Lund University, Box 118, SE-221 00 Lund, Sweden}
\newcommand{\lyon}{IPNL, CNRS/IN2P3, Univ Lyon, Université Lyon 1, F-69622, Villeurbanne, France}
\newcommand{\maryland}{University of Maryland, College Park, Maryland 20742, USA}
\newcommand{\michigan}{Department of Physics, University of Michigan, Ann Arbor, Michigan 48109-1040, USA}
\newcommand{\muhlenberg}{Muhlenberg College, Allentown, Pennsylvania 18104-5586, USA}
\newcommand{\nara}{Nara Women's University, Kita-uoya Nishi-machi Nara 630-8506, Japan}
\newcommand{\natmephi}{National Research Nuclear University, MEPhI, Moscow Engineering Physics Institute, Moscow, 115409, Russia}
\newcommand{\newmex}{University of New Mexico, Albuquerque, New Mexico 87131, USA}
\newcommand{\nmsu}{New Mexico State University, Las Cruces, New Mexico 88003, USA}
\newcommand{\ohio}{Department of Physics and Astronomy, Ohio University, Athens, Ohio 45701, USA}
\newcommand{\ornl}{Oak Ridge National Laboratory, Oak Ridge, Tennessee 37831, USA}
\newcommand{\orsay}{IPN-Orsay, Univ.~Paris-Sud, CNRS/IN2P3, Universit\'e Paris-Saclay, BP1, F-91406, Orsay, France}
\newcommand{\peking}{Peking University, Beijing 100871, People's Republic of China}
\newcommand{\pnpi}{PNPI, Petersburg Nuclear Physics Institute, Gatchina, Leningrad region, 188300, Russia}
\newcommand{\riken}{RIKEN Nishina Center for Accelerator-Based Science, Wako, Saitama 351-0198, Japan}
\newcommand{\rikjrbrc}{RIKEN BNL Research Center, Brookhaven National Laboratory, Upton, New York 11973-5000, USA}
\newcommand{\rikkyo}{Physics Department, Rikkyo University, 3-34-1 Nishi-Ikebukuro, Toshima, Tokyo 171-8501, Japan}
\newcommand{\saispbstu}{Saint Petersburg State Polytechnic University, St.~Petersburg, 195251 Russia}
\newcommand{\seoulnat}{Department of Physics and Astronomy, Seoul National University, Seoul 151-742, Korea}
\newcommand{\stonycrkp}{Department of Physics and Astronomy, Stony Brook University, SUNY, Stony Brook, New York 11794-3800, USA}
\newcommand{\tenn}{University of Tennessee, Knoxville, Tennessee 37996, USA}
\newcommand{\titech}{Department of Physics, Tokyo Institute of Technology, Oh-okayama, Meguro, Tokyo 152-8551, Japan}
\newcommand{\tsukuba}{Center for Integrated Research in Fundamental Science and Engineering, University of Tsukuba, Tsukuba, Ibaraki 305, Japan}
\newcommand{\vandy}{Vanderbilt University, Nashville, Tennessee 37235, USA}
\newcommand{\weizmann}{Weizmann Institute, Rehovot 76100, Israel}
\newcommand{\wigner}{Institute for Particle and Nuclear Physics, Wigner Research Centre for Physics, Hungarian Academy of Sciences (Wigner RCP, RMKI) H-1525 Budapest 114, POBox 49, Budapest, Hungary}
\newcommand{\yonsei}{Yonsei University, IPAP, Seoul 120-749, Korea}
\newcommand{\zagreb}{Department of Physics, Faculty of Science, University of Zagreb, Bijeni\v{c}ka c.~32 HR-10002 Zagreb, Croatia}
\affiliation{\abilene}
\affiliation{\augie}
\affiliation{\banaras}
\affiliation{\barc}
\affiliation{\baruch}
\affiliation{\bnlcoll}
\affiliation{\bnlphys}
\affiliation{\caucr}
\affiliation{\charlesczech}
\affiliation{\chonbuk}
\affiliation{\cns}
\affiliation{\colorado}
\affiliation{\columbia}
\affiliation{\czechtech}
\affiliation{\debrecen}
\affiliation{\elte}
\affiliation{\eszterhazy}
\affiliation{\ewha}
\affiliation{\fsu}
\affiliation{\gsu}
\affiliation{\hiroshima}
\affiliation{\howard}
\affiliation{\ihepprot}
\affiliation{\illuiuc}
\affiliation{\inrras}
\affiliation{\instpasczech}
\affiliation{\isu}
\affiliation{\jaea}
\affiliation{\jyvaskyla}
\affiliation{\kek}
\affiliation{\korea}
\affiliation{\kurchatov}
\affiliation{\kyoto}
\affiliation{\lawllnl}
\affiliation{\losalamos}
\affiliation{\lund}
\affiliation{\lyon}
\affiliation{\maryland}
\affiliation{\michigan}
\affiliation{\muhlenberg}
\affiliation{\nara}
\affiliation{\natmephi}
\affiliation{\newmex}
\affiliation{\nmsu}
\affiliation{\ohio}
\affiliation{\ornl}
\affiliation{\orsay}
\affiliation{\peking}
\affiliation{\pnpi}
\affiliation{\riken}
\affiliation{\rikjrbrc}
\affiliation{\rikkyo}
\affiliation{\saispbstu}
\affiliation{\seoulnat}
\affiliation{\stonycrkp}
\affiliation{\tenn}
\affiliation{\titech}
\affiliation{\tsukuba}
\affiliation{\vandy}
\affiliation{\weizmann}
\affiliation{\wigner}
\affiliation{\yonsei}
\affiliation{\zagreb}
\author{C.~Aidala} \affiliation{\michigan} 
\author{Y.~Akiba} \email[PHENIX Spokesperson: ]{akiba@rcf.rhic.bnl.gov} \affiliation{\riken} \affiliation{\rikjrbrc} 
\author{M.~Alfred} \affiliation{\howard} 
\author{K.~Aoki} \affiliation{\kek} 
\author{N.~Apadula} \affiliation{\isu} 
\author{C.~Ayuso} \affiliation{\michigan} 
\author{V.~Babintsev} \affiliation{\ihepprot} 
\author{A.~Bagoly} \affiliation{\elte} 
\author{K.N.~Barish} \affiliation{\caucr} 
\author{S.~Bathe} \affiliation{\baruch} \affiliation{\rikjrbrc} 
% ...
\collaboration{PHENIX Collaboration} \noaffiliation

\date{\today}


\begin{abstract}


We present measurements of the elliptic flow ($v_2$) as a function of 
transverse momentum ($p_T$), pseudorapidity ($\eta$), and centrality in 
$d$$+$Au collisions at $\sqrt{s_{_{NN}}}=200$, 62.4, 39, and 19.6 GeV. 
The beam-energy scan of $d$$+$Au collisions provides a testing ground 
for the onset of flow signatures in small collision systems.  We measure 
a nonzero $v_2$ signal at all four collision energies, which, at 
midrapidity and low $p_T$, is consistent with predictions from viscous 
hydrodynamic models. Comparisons with calculations from parton transport 
models (based on the {\sc ampt} Monte Carlo generator) show good 
agreement with the data at midrapidity to forward ($d$-going) rapidities 
and low $p_T$. At backward (Au-going) rapidities and $p_T>1.5$ GeV/$c$, 
the data diverges from {\sc ampt} calculations of $v_2$ relative to the 
initial geometry, indicating the possible dominance of nongeometry 
related correlations, referred to as nonflow.  We also present 
measurements of the charged-particle multiplicity ($dN_{\rm ch}/d\eta$) 
as a function of $\eta$ in central $d$$+$Au collisions at the same 
energies.  We find that in $d$$+$Au collisions at $\sqrt{s_{_{NN}}}=200$ 
GeV the $v_2$ scales with $dN_{\rm ch}/d\eta$ over all $\eta$ in the 
PHENIX acceptance. At $\sqrt{s_{_{NN}}}=62.4$, and 39 GeV, $v_2$ scales 
with $dN_{\rm ch}/d\eta$ at midrapidity and forward rapidity, but falls 
off at backward rapidity. This departure from the $dN_{\rm ch}/d\eta$ 
scaling may be a further indication of nonflow effects dominating at 
backward rapidity.

\end{abstract}

\maketitle

\section{Introduction}
\label{sec:introduction}

Measurements of the azimuthal momentum anisotropy of particles produced 
in high-energy heavy ion collisions ($A$$+$$A$) have provided strong 
evidence for the formation of a strongly coupled Quark-Gluon Plasma 
(QGP)\cite{Arsene:2004fa,Adcox:2004mh,Back:2004je,Adams:2005dq}. This 
anisotropy, as measured by the Fourier coefficients, $v_n$, can be 
understood as arising from initial geometry propagated to final-state 
momentum correlations via interactions between medium constituents. 
These interactions have been well described by relativistic 
hydrodynamics with a low ratio of viscosity to entropy 
density~\cite{Romatschke:2009im,Heinz:2013th}.

In 2012, measurements of $v_2$ in \sqsn~=~5.02 TeV \ppb collisions at 
the Large Hadron Collider 
(LHC)~\cite{Aad:2012gla,Abelev:2012ola,CMS:2012qk} and \sqsn~=~200 GeV 
\dau collisions at the Relativistic Heavy Ion Collider 
(RHIC)~\cite{Adare:2013piz} raised the question whether a QGP might be 
formed even in these small collision systems. Further measurements in 
\ppb collisions revealed that the signal persists for multi-particle 
correlations~\cite{Aad:2013fja,Chatrchyan:2013nka,Abelev:2014mda,Khachatryan:2015waa}, 
which is additional evidence of collective behavior. To test the 
signal's connection to the initial geometry of the collision, PHENIX 
measured $v_2$ in \pdheau collisions and $v_3$ in \heau collisions at 
\sqsn~=~200 
GeV~\cite{Adare:2014keg,Adare:2015ctn,Aidala:2016vgl,Nagle:2013lja}. The 
results are consistent with the interpretation that the measured $v_2$ 
arises from initial geometry. High-multiplicity \pp collisions at 
\sqsn~=~2.76, 5.02, 7.13, and 13 TeV exhibit similar 
effects~\cite{Aad:2015gqa,Khachatryan:2010gv,Khachatryan:2016txc} and 
may also be related to the initial geometry~\cite{Weller:2017tsr}.

Even in these small collision systems, the data at both RHIC and the LHC 
can be described by hydrodynamic 
calculations~\cite{Weller:2017tsr,Aidala:2016vgl}. However, it has also 
been shown that calculations using kinetic theories of hadronic and 
partonic scattering (e.g., a multiphase transport (\ampt) 
model~\cite{Lin:2004en}) can qualitatively describe the $v_2$ measured 
in small systems~\cite{Bozek:2015swa,Aidala:2016vgl,Koop:2015wea}. In 
both hydrodynamic and kinetic models, initial geometry (coordinate space 
anisotropy) is translated to final state momentum space anisotropy via 
interactions between medium constituents. In contrast, other 
explanations, including color recombination~\cite{Ortiz:2013yxa} and 
initial-state effects from glasma diagrams~\cite{Dusling:2012iga}, have 
also been proposed, where the final-state momentum correlations are due 
to initial momentum correlations rather than a connection to the initial 
geometry.

Throughout this paper we use a working definition of ``flow'' as initial 
geometry propagated to final-state azimuthal momentum anisotropy, 
regardless of the mechanism of propagation (e.g. fluid flow or particle 
transport). All other sources of final-state azimuthal momentum 
anisotropy are referred to as ``nonflow''. Examples of nonflow include 
jet correlations, resonance decays, and Coulomb interactions.

In 2016, RHIC delivered \dau collisions at \sqsn~=~200, 62.4, 39, and 
19.6 GeV in order to investigate the onset of collectivity. PHENIX has 
previously published results on multi-particle correlations from this 
data set~\cite{Aidala:2017ajz}, providing evidence for collective behavior at 
all energies. Here we report comprehensive measurements of $v_2$ as a 
function of \pt, $\eta$, and centrality in \dau collisions at 
\sqsn~=~200, 62.4, 39, and 19.6 GeV. We also report measurements of the 
charged particle multiplicity (\dndeta) as a function of $\eta$ in 
central \dau collisions at the same energies.

\section{Experiment and Data Set}
\label{sec:experiment}

\begin{figure}[htbp]
    \includegraphics[width=1.0\linewidth]{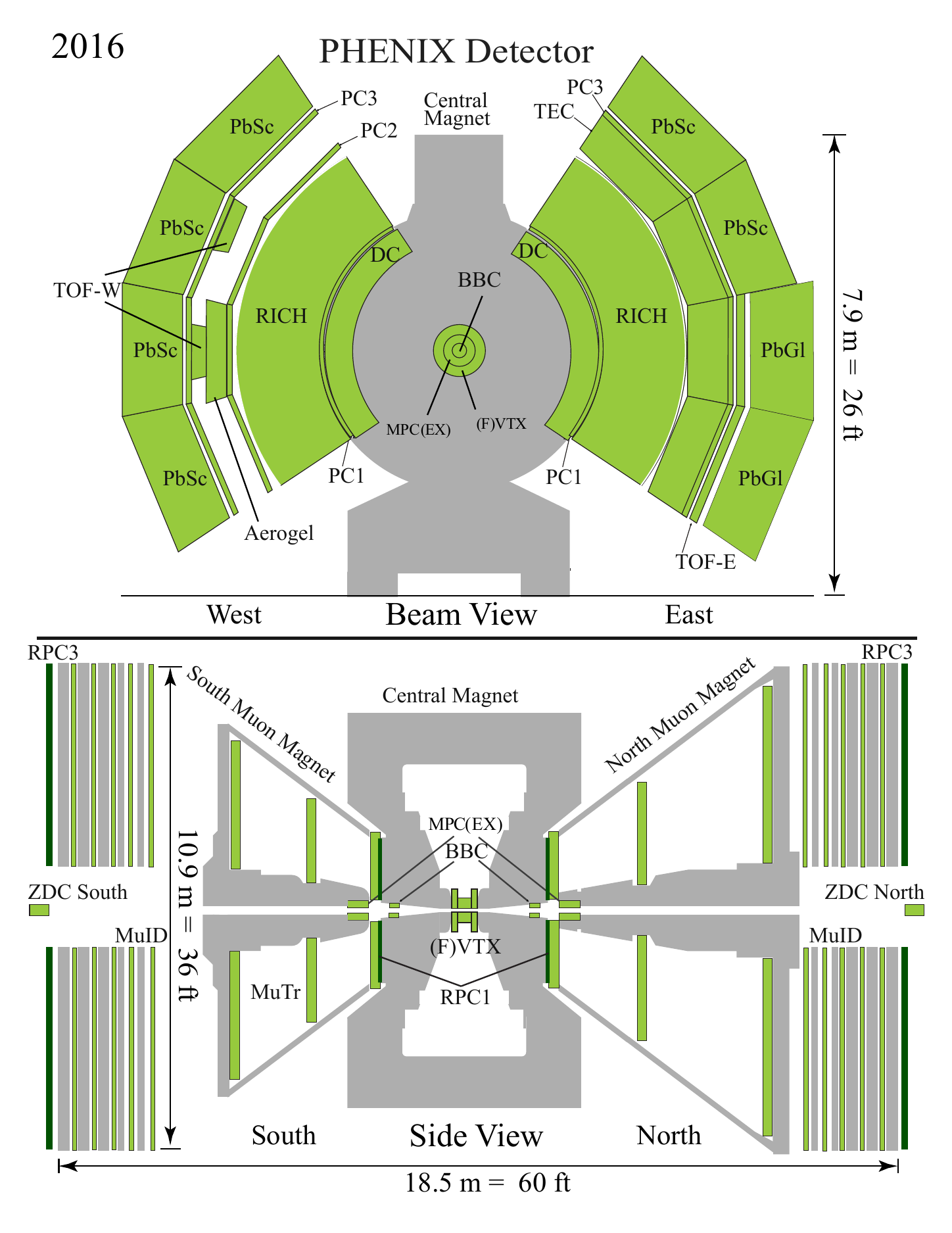}
    \caption{\label{fig:phenix} 
A schematic view of the PHENIX detector as configured in 2016.}
\end{figure}

The PHENIX detector is described in detail in Ref.~\cite{Adcox:2003zm} 
and shown schematically in Fig.~\ref{fig:phenix}. Global event 
characterization and triggering use two beam-beam counters 
(BBC)\cite{Allen:2003zt} located in the pseudorapidity region 
$3.1<|\eta|<3.9$, as well as a forward silicon vertex detector 
(FVTX)~\cite{Aidala:2013vna} covering $1<|\eta|<3$. Each BBC comprises 
64 \v{C}erenkov counters arrayed around the beam pipe 1.44 m from the 
nominal interaction region.  The counters comprise 3 cm of quartz coupled 
to a mesh-dynode photomultiplier tube, where the charge is calibrated to 
a minimum-ionizing charged particle.  The FVTX is made up of two annular 
endcaps, each with four stations of silicon mini-strip sensors.  Each 
station comprises 47 individual silicon sensors, each of which contains 
two columns of mini-strips with 75 $\mu$m pitch in the radial direction 
and lengths in the $\phi$ direction varying from 3.4 mm at the inner 
radius to 11.5 mm at the outer radius.  The negative-rapidity 
$south$-side region (Au-going direction) has the BBCS and FVTXS arms, 
while the positive-rapidity $north$-side region ($d$-going direction) 
has the BBCN and FVTXN arms.  Charged-particle tracking is provided by
the $east$ and $west$ central arms at midrapidity, covering 
$|\eta|<0.35$ each with an azimuthal ($\phi$) coverage of $\pi/2$.

At 200 and 62.4 GeV a minimum bias (MB) interaction trigger is provided 
by the BBC.  For the MB trigger, at least one hit tube is required in each 
of the north and south detectors.  The fraction of the \dau inelastic 
cross section that the MB trigger fires on, $\epsilon_{\rm MB}$, is given in 
Table~\ref{tab:data} for both energies. In addition to the MB trigger, a 
high-multiplicity trigger that required $>40$ (29) hit tubes in the BBCS 
for 200 (62.4) GeV was also run, providing a factor of 188 (11) 
enhancement of high-multiplicity events. Analyzed events were further 
required to have a reconstructed collision vertex in the longitudinal 
direction as reconstructed by the BBC of $|z_{\rm vrtx}|<10$ cm. The 
resulting number of analyzed events is shown in Table~\ref{tab:data}.

\begin{table}
    \caption{\label{tab:data}
Summary of the data analyzed by PHENIX from the 2016 RHIC \dau beam 
energy scan.}
    \begin{ruledtabular} \begin{tabular}{cccc}
         &                     & \# Analyzed      & \# Analyzed \\
         &                     & MB               & high-multiplicity \\
  \sqsn [GeV] & $\epsilon_{\rm MB}$ & triggered        & triggered \\
              &                     & events [$10^6$]  & events [$10^6$] \\
    \hline
    200  & 88$\pm$4\% &   53 & 569 (0\%--5\%) \\
    62.4 & 78$\pm$4\% &  113 & 214 (0\%--10\%) \\
    39   & 74$\pm$6\% &  231 & 171 (0\%--20\%) \\
    19.6 & 61$\pm$8\% &   33 &   7 (0\%--20\%) \\
	\end{tabular} \end{ruledtabular} 
\end{table}

At 39 and 19.6 GeV, the FVTX combined with the south BBC is used for the 
MB trigger. This combination has a larger trigger efficiency at these 
lower energies than a BBC coincidence due to the low multiplicities in 
the region $3.1<\eta<3.9$ at these energies. The FVTX trigger requires 
at least one hit in 3 of the 4 stations of the FVTX in a given sector 
covering approximately $\Delta\phi=0.26$ rad, effectively requiring a 
single track in each of the north and south arms. To reduce background, 
at least one hit tube was required in the south BBC. The efficiency of 
the MB trigger, $\epsilon_{\rm MB}$ at both energies is given in 
Table~\ref{tab:data}. Additionally, a high-multiplicity trigger was 
implemented that further required $>27$ (18) hits in the south BBC for 
39 (19.6) GeV, providing a factor of 6.0 (1.8) enhancement of 
high-multiplicity events. Analyzed events were also required to have 
$|z_{\rm vrtx}|<10$ cm, as reconstructed by the FVTX. To reduce beam-gas and 
beam-pipe background, the total number of reconstructed clusters in the 
FVTX, both south and north arms, was required to be $<$~500 (300) at 39 
(19.6) GeV. The resulting number of analyzed events is shown in 
Table~\ref{tab:data}.

\begin{figure}[htbp]
	\includegraphics[width=0.95\linewidth]{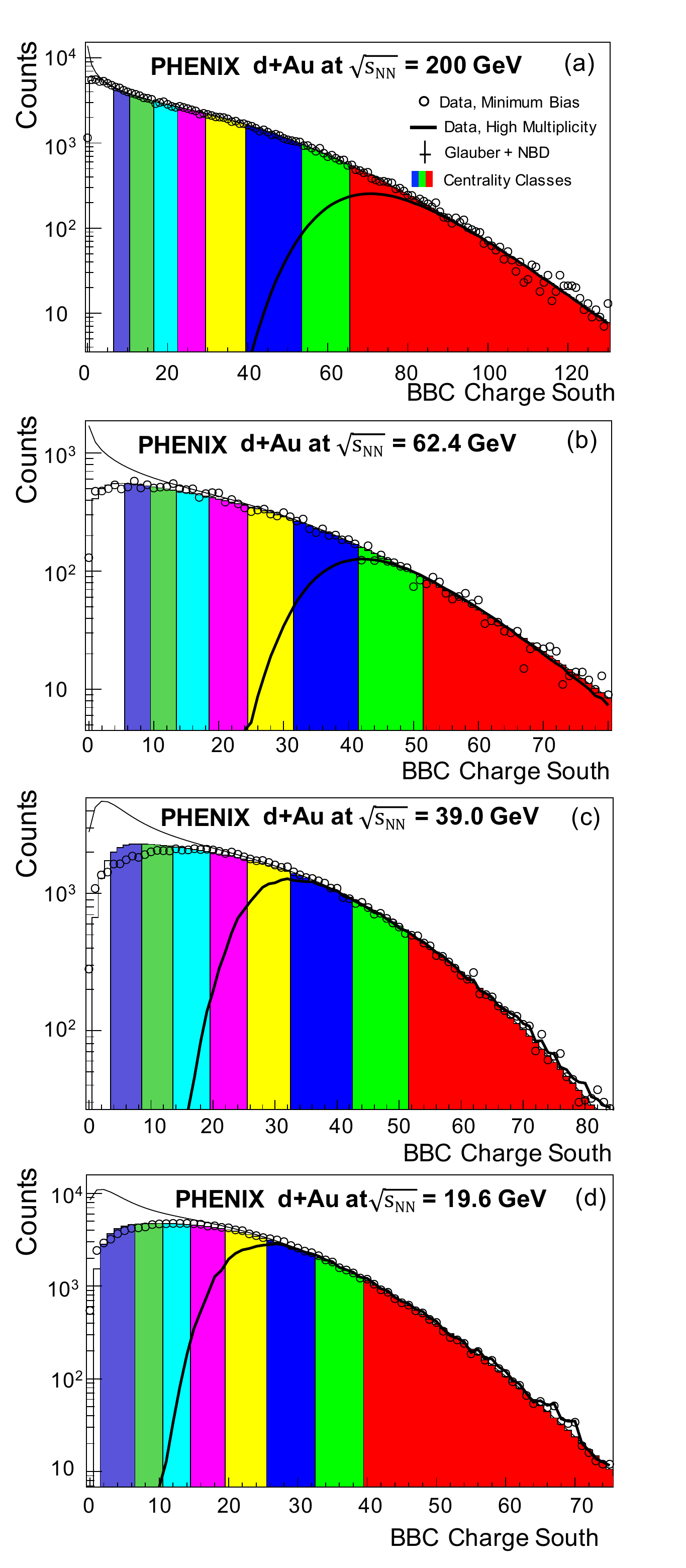}
	\caption{\label{fig:qbbcs} 
The distributions of total charge in the BBCS for \dau collisions at 
\sqsn~=~200 (a), 62.4 (b), 39 (c), and 19.6 (d) GeV. The data is from MB 
collisions, and we note that, as discussed in the text, the MB trigger 
definition changes with energy. The colored bands represent, from right 
to left, the centrality categorizations 0\%--5\%, 5\%--10\%, 10\%--15\%, 
15\%--20\%, 20\%--30\%, 30\%--40\%, 40\%--50\%, 50\%--60\%, and 
60\%--XX\%, where XX is the value of the MB trigger efficiency for each 
energy, given in Table~\ref{tab:data}. The thick solid line shows the 
high-multiplicity trigger selection, scaled down to match the MB 
distribution.}
\end{figure}

The collision centrality at all four energies is determined using the 
total charge in the south (Au-going) BBC, as described in 
Ref.~\cite{Adare:2013nff}. Figure~\ref{fig:qbbcs} shows the BBCS charge 
distributions from MB triggered data at each energy along with the 
limits of the various centrality bins. It also includes the BBCS charge 
distributions for the high-multiplicity trigger, renormalized to match 
the high-charge region, showing the trigger turn-on at each energy. To 
avoid bias in the centrality distribution, analyzed events firing the 
high-multiplicity trigger are required to have centrality 0\%--5\%, 
0\%--10\%, 0\%--20\%, 0\%--20\% at \sqsn~=~200, 62.4, 39, and 19.6, 
respectively. These regions correspond to centralities for which the 
high-multiplicity trigger was efficient.

Using Monte-Carlo Glauber combined with fluctuations modeled by a 
negative binomial distribution as laid out in Ref.~\cite{Adare:2013nff}, 
the mean number of participants, $\langle \Npart \rangle$, and the mean 
initial geometry eccentricity, $\langle \varepsilon_{2} \rangle$, can be 
characterized for given centrality bins. Table~\ref{tab:cent} shows the 
$\langle \Npart \rangle$ and $\langle \varepsilon_{2} \rangle$ values 
for central collisions at all four energies. The $\langle 
\varepsilon_{2} \rangle$ values are consistent at all four collision 
energies within uncertainties. The $\langle \Npart \rangle$ values, 
however, decrease with decreasing energy. This can be attributed to both 
the decreasing nucleon-nucleon interaction cross section and the larger 
centrality bins at 39 and 19.6 GeV, which were used to improve the 
statistical precision of the measurements.

\begin{table}
	\caption{\label{tab:cent} 
Summary of the mean number of participants, $\langle \Npart \rangle$, 
and eccentricity, $\langle \varepsilon_{2} \rangle$, for central \dau 
collisions at \sqsn~=~200, 62.4, 39, and 19.6 GeV.}
    \begin{ruledtabular} \begin{tabular}{cccc}
	\sqsn [GeV] 
& centrality & $\langle \Npart \rangle$ & $\langle \varepsilon_{2} \rangle$\\
	\hline
	200 & 0\%--5\% & 17.8$\pm$1.2 & 0.54$\pm$0.04 \\
	62.4 & 0\%--5\% & 16.3$\pm$1.0 & 0.55$\pm$0.05 \\
	39 & 0\%--10\% & 15.9$\pm$1.0 & 0.56$\pm$0.06 \\
	19.6 & 0\%--20\% & 13.6$\pm$1.0 & 0.55$\pm$0.05 \\
	\end{tabular} \end{ruledtabular}
\end{table}

In the central arms, unidentified charged particle tracking uses the 
drift chamber (DC) and pad chamber (PC) layers. We require tracks to 
have a unique match between DC hits and PC hits in the layer immediately 
surrounding the DC. Tracks are further required to have a matching hit 
in the third PC layer at $R=4.98$ m that is within $\pm3\sigma$ of the 
projected track location, where $\sigma$ characterizes the 
momentum-dependent widths of the matching distributions.

In addition to triggering, the FVTX is used for unidentified charged 
particle tracking. The FVTX does not measure track momentum, and we 
therefore are limited to a momentum integrated measurement. We require 
reconstructed tracks in the FVTX to have hits in at least 3 of the 4 
stations with fit quality, $\chi^2/$d.o.f.$<5$. We further require that 
the distance of closest approach of the track to the primary collision 
vertex, $DCA$, be within 2.0 cm in both the $x$ and $y$ directions, 
transverse to the beam axis. The expected $DCA$ resolution from 
simulation is $\approx1.2$ cm at 500 MeV. This loose cut on the $DCA$ 
removes background from upstream beam-gas interactions, as well as 
mis-reconstructed tracks.

The luminosity delivered by RHIC for \dau collisions at \sqsn~=~200 GeV 
is high enough that approximately 6\% of events are expected to contain 
multiple collisions (i.e. pile-up). The fraction of pile-up events is 
larger in central events, and is expected to be as large as 20\% in the 
highest luminosity periods. An algorithm was developed to aid in 
rejecting these events. For each event, the distribution of times for 
each hit tube in the BBCS is determined. Then, the fraction, $f$, of the 
time distribution for that event which is within a 0.5 ns window of the 
mode of the measured distribution is calculated. Because multiple collisions 
typically occur at different positions along the beam axis, particles 
from these collisions tend to leave multiple peaks in the distribution 
of times recorded in the BBCS. Therefore, pile-up events are typically 
characterized by low values of $f$. We reject events with $f<0.95$ for 
centrality 0\%--20\%. Studies using low luminosity data and manufactured 
pile-up events indicate that this cut rejects 81\% of pile-up events 
while accepting 93\% of single collision events for 0\%--5\% central 
collisions. Based on the luminosities delivered at 62.4, 39, and 19.6 
GeV, fewer than 1\% of events are expected to contain multiple 
collisions, and therefore no cut on $f$ is included.

\section{Analysis}
\label{sec:analysis}

We first discuss two-particle correlation functions in 
Sec.~\ref{sec:2pcorr}. The analysis of the \pt dependence of the second 
order flow coefficient, $v_2$, is discussed in Sec.~\ref{sec:v2pt}. The 
analysis of the $\eta$ dependence of $v_2$ is discussed in 
Sec.~\ref{sec:v2eta}. The analysis of \dndeta is discussed in 
Sec.~\ref{sec:dndeta}.

\subsection{Two-particle correlations}
\label{sec:2pcorr}

\begin{figure*}[htbp]
	\includegraphics[width=0.99\textwidth]{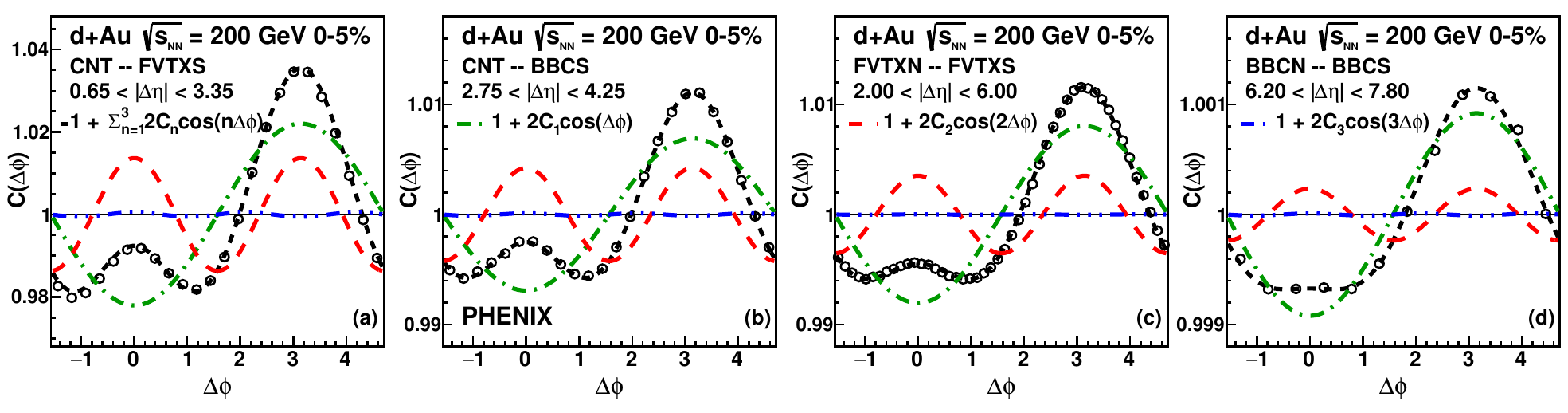}
	\caption{\label{fig:2pcorrdeta} 
Two-particle $\Delta\phi$ correlations in central \dau collisions at 
\sqsn~=~200 GeV between various detectors. The blue dot-dashed lines, 
red long-dashed lines, and green dotted lines, correspond to the $C_1$, 
$C_2$, and $C_3$ components, respectively. The black dashed lines 
correspond to the sum of the $C_n$'s up to third order. For the 
correlations in panels (a) and (b), CNT tracks were required to be 
within $0.2<\pt\ [{\rm GeV}/c]< 5.0$.}
\end{figure*}

We start by constructing long-range azimuthal correlations in \dau 
collisions at \sqsn~=~200 GeV. The two-particle correlation function is 
defined as

\begin{equation}
	C(\Delta\phi) = \frac{S(\Delta\phi)}{M(\Delta\phi)}\frac{\int_0^{2\pi}M(\Delta\phi)}{\int_0^{2\pi}S(\Delta\phi)},
\end{equation}
where $\Delta\phi$ is the difference in the azimuthal angles between two 
tracks, $S(\Delta\phi)$ is the signal distribution, constructed from 
track pairs in the same event, and $M(\Delta\phi)$ is the mixed event 
distribution, constructed from track pairs from different events in the 
same centrality and collision vertex class. Figure~\ref{fig:2pcorrdeta} 
shows $C(\Delta\phi)$ for correlations of tracks between different 
detectors in central \dau collisions at \sqsn~=~200 GeV: (a) between 
tracks in the central arms and tracks in the FVTXS, (b) between tracks 
in the central arms and tubes in the BBCS, (c) between tracks in the 
FVTXS and FVTXN, and (d) between tubes in the BBCS and BBCN. By 
comparing $C(\Delta\phi)$ distributions between different sets of 
detectors we naturally change the $\Delta\eta$ requirement for the pair 
of tracks. Correlations with a small $\Delta\eta$ are typically thought 
to be dominated by nonflow correlations, particularly from intrajet 
correlations near $\Delta\phi=0$, as well as dijet correlations near 
$\Delta\phi=\pi$. By increasing the $\Delta\eta$ gap between particles 
we naturally reduce the dominance of these nonflow correlations. 
Figure~\ref{fig:2pcorrdeta} shows correlation functions with (a) 
$0.65<|\Delta\eta|<3.35$, (b) $2.75<|\Delta\eta|<4.25$, (c) 
$2.0<|\Delta\eta|<6.0$, and (d) $6.2<|\Delta\eta|<7.8$.

The correlations exhibit two visible peaks at $\Delta\phi=0$ and 
$\Delta\phi=\pi$. The peak at $\Delta\phi=\pi$ is associated with, for 
example, dijets. The peak at $\Delta\phi=0$ does not arise from 
particles within a jet or decays, because we have imposed a large 
$\Delta\eta$ gap. This peak was first observed in $A$$+$$A$ collisions and 
has been termed the long-range near-side ridge. This near-side ridge was 
one of the key components in understanding the hydrodynamic description 
of $A$$+$$A$ collisions (See Ref.~\cite{Sorensen:2011hm} and references 
therein). The observation of this structure in high-multiplicity \pp 
collisions at \sqsn~=~7 TeV~\cite{Khachatryan:2010gv} was one of the 
first hints that collectivity may exist even in small collision systems. 
We observe a visible near-side ridge up to $|\Delta\eta|>2.75$.

To investigate these correlations further, we fit the 
distribution with a Fourier series up to 3rd order:
	\begin{equation}
	F(\Delta\phi) = 1 + \sum_{n=1}^3 2 C_n \cos{n\Delta\phi},
	\end{equation}
where $C_n$ is the nth order Fourier component. The full fit and the 
components are shown as lines in Fig.~\ref{fig:2pcorrdeta}. The dominant 
term is the first order $C_1$ term, and arises from elementary 
processes, such as momentum conservation. The second order term, $C_2$, 
is associated with flow. While the longest range correlation shown in 
Fig.~\ref{fig:2pcorrdeta}(d), with $|\Delta\eta|>6.2$, does not show a 
clear peak at $\Delta\phi=0$, it does include a strong second-order 
Fourier component, $C_2$.

\begin{figure}[htbp]
	\includegraphics[width=1.0\linewidth]{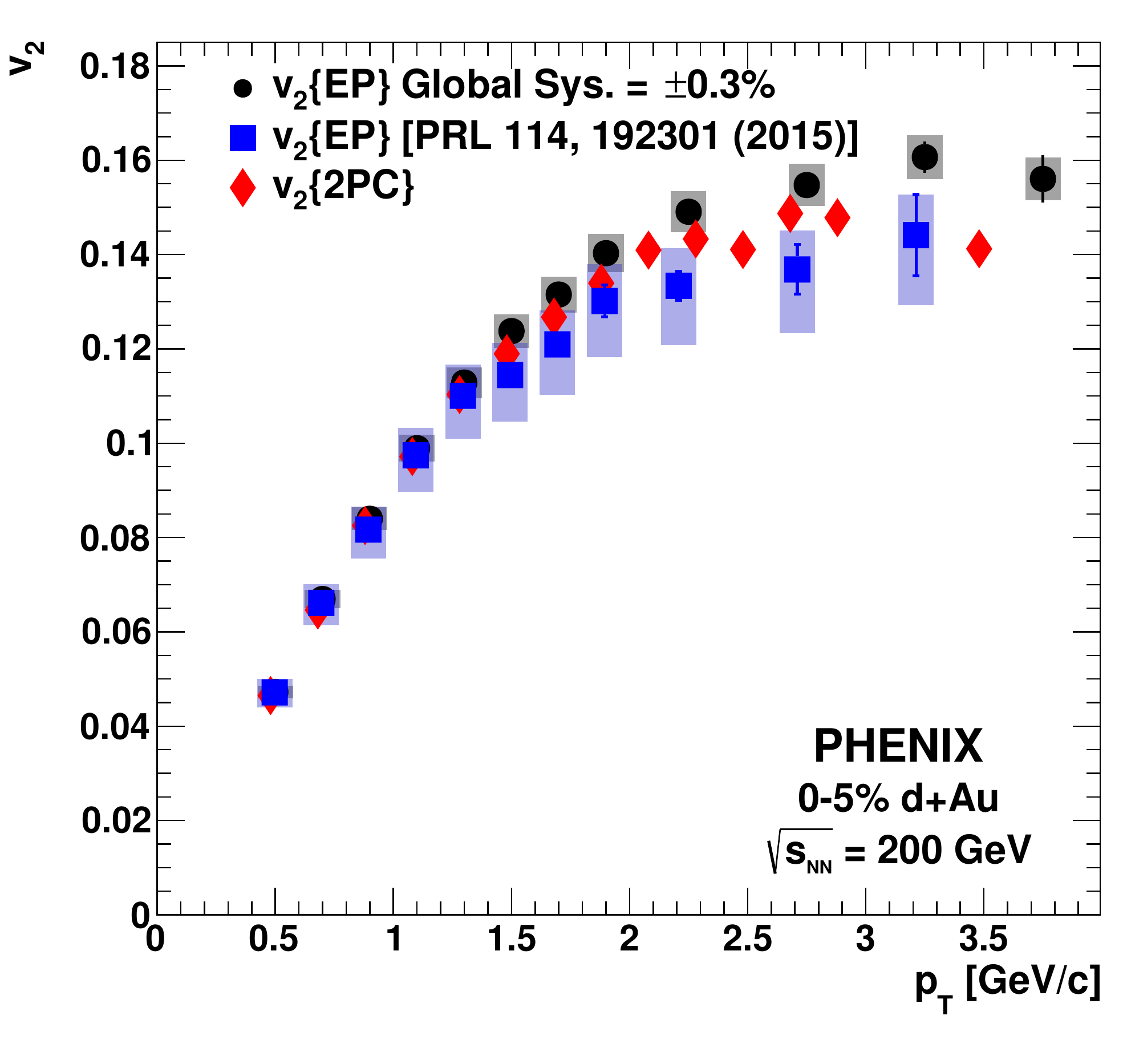}
	\caption{\label{fig:v2meth} 
The $v_2$ vs \pt in 0\%--5\% central \dau collisions at \sqsn~=~200 GeV 
using the event-plane method (black filled circles) and two-particle 
correlations (red filled diamonds). Also shown are the previously 
published $v_2$ vs \pt using the event-plane method (blue filled 
squares) from PHENIX using data collected in 2008~\cite{Adare:2014keg}.}
\end{figure}

Using the two-particle correlation (2PC) functions $C(\Delta\phi, \pt)$, the 
$v_2$ as a function of \pt, $v_2\{2PC\}$, can be calculated for central 
arm tracks using

\begin{equation}
v_2\{2PC\} = \sqrt{\frac{C^{AB}(\Delta\phi, \pt)\times C^{AC}(\Delta\phi, \pt)}{C^{BC}(\Delta\phi)}},
\label{eq:2pc_abc}
\end{equation}
where the superscript $AB$ refers to correlations between central arm 
and FVTXS tracks, $AC$ refers to correlations between central arm tracks 
and BBCS tubes, and $BC$ refers to correlations between FVTXS tracks and 
BBCS tubes. This relation can be understood as arising from the 
assumption of flow factorization, which allows the correlation function 
to be interpreted as e.g. $C^{AB}(\Delta\phi) = \langle 
v_n^Av_n^B\rangle$. In that way, Eqn.~\ref{eq:2pc_abc} reduces to
\begin{equation}
v_2\{2PC\} = \sqrt{\frac{\langle v_n^Av_n^B\rangle\langle v_n^Av_n^C\rangle}{\langle v_n^Bv_n^C\rangle}},
\label{eq:2pc_abc_vn}
\end{equation}
where the superscripts $A$, $B$, $C$ represent the central arms, the 
FVTXS, and the BBCS, respectively.

The $v_2\{2PC\}$ vs \pt for 0\%--5\% \dau collisions at \sqsn~=~200 GeV 
is shown as the red points in Fig.~\ref{fig:v2meth}.

\begin{figure*}[htbp]
	\includegraphics[width=0.99\textwidth]{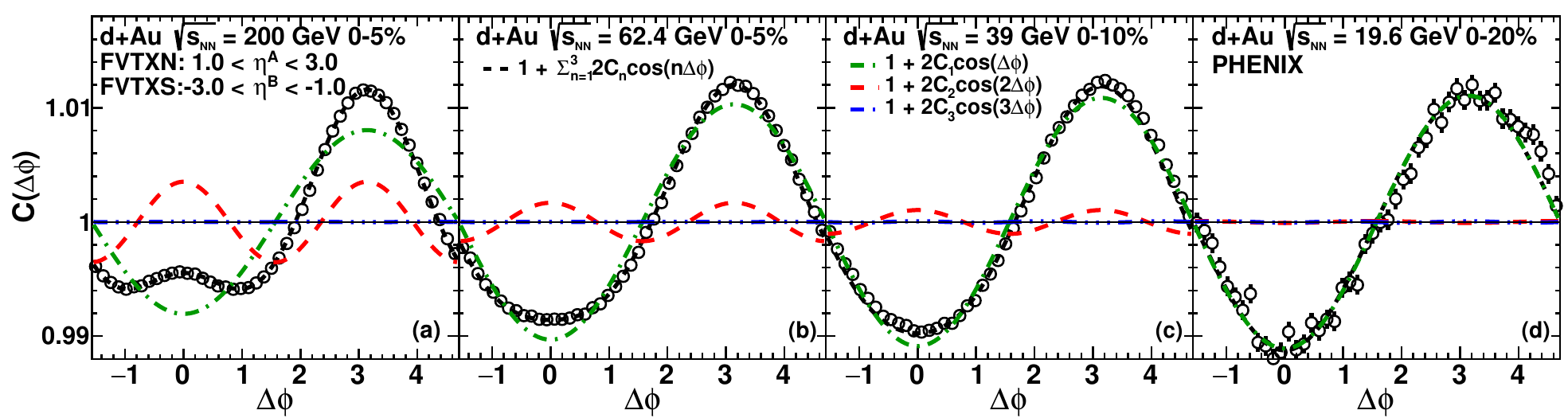}
	\caption{\label{fig:2pcorrfvtx} 
Two-particle $\Delta\phi$ correlations in central \dau collisions at 
\sqsn~=~200, 62.4, 39, and 19.6 GeV between tracks in the north and 
south FVTX detectors. The blue dot-dashed lines, red long-dashed lines, 
and green dotted lines, correspond to the $C_1$, $C_2$, and $C_3$ 
components, respectively. The black dashed lines correspond to the sum 
of the $C_n$'s up to third order.}
\end{figure*}

We also investigate the energy dependence of the near-side ridge using 
correlations between tracks in the FVTXN and FVTXS. 
Figure~\ref{fig:2pcorrfvtx} shows $C(\Delta\phi)$ with 
$2.0<|\Delta\eta|<6.0$ for central \dau collisions at \sqsn~=~200, 62.4, 
39, and 19.6 GeV. A visible peak at $\Delta\phi=0$ is only observed at 
200 GeV; however, substantial $C_2$ components are extracted at 62.4 and 
39 GeV. At 19.6 GeV, no visible $C_2$ component is extracted. The 
$C(\Delta\phi)$ is integrated over \pt and hence dominated by low \pt 
tracks. Therefore, the lack of a visible $C_2$ component at 19.6 GeV 
does not exclude a nonzero $v_2$, particularly at higher \pt.

\subsection{Analysis of $v_2$ vs \pt using the event-plane method}
\label{sec:v2pt}

The standard event-plane method~\cite{Poskanzer:1998yz} is used to 
calculate $v_2$ as a function of \pt:
\begin{equation}
	v_2(\pt) = \frac{\langle\cos{2(\phi_{\rm trk}(\pt) - \psifvtxs)}\rangle}{R(\psifvtxs)},
	\label{eq:v2pt}
\end{equation}

where $\phi_{\rm trk}$ is the azimuthal angle of tracks in the central arms, 
and $\psifvtxs$ is the azimuthal angle of the second-order event-plane 
measured by the FVTXS. The event plane in the FVTXS is constructed in 
the usual way of $2\psi_2 = {\rm atan2}(Q_{y}^{\rm FVTXS},Q_{x}^{\rm 
FVTXS})$, with $Q^{\rm FVTXS} = \sum_{i=1}^M e^{in\phi_i}$, where 
$\phi_i$ is the azimuthal angle of some cluster in the FVTXS. The 
underlying physics correlation is the same whether one uses tracks or 
clusters, but the use of clusters provides higher event-plane resolution 
and therefore greater statistical precision. The resolution of 
$\psifvtxs$, $R(\psifvtxs)$, is calculated using the three-subevent 
method~\cite{Poskanzer:1998yz} that correlates measurements in the 
FVTXS, BBCS, and central arms. The resolution is strongly dependent on 
both the collision energy and centrality, and is shown in 
Table~\ref{tab:respsi2}. We note that for 39 to 200 GeV we find that 
$R(\psifvtxs)$ increases in the most peripheral centrality bin. 
This is contrary to expectations, because $R(\psifvtxs)$ depends on both 
the $v_2$ in the event-plane region and the number of particles, both of 
which are expected to decrease in more peripheral events. Nonflow is 
likely the largest contribution in the most peripheral collisions, and 
may result in this increased resolution.

Due to its better resolution, we use the measurement of $\Psi_2$ from 
the FVTXS. However, we can compare the $v_2$ vs \pt measured using the 
BBCS, which has a larger separation of $|\Delta\eta|>2.75$ relative to 
the central arm tracks compared to $|\Delta\eta|>0.65$ with the FVTXS. 
The $v_2$ values are found to agree within 2.5\% for $\pt<2$ GeV/$c$, 
where we expect nonflow effects to be small. For $\pt>2$ GeV/$c$ a 
larger value of $v_2$ is observed using the FVTXS compared to the BBCS. 
This difference is likely due to differences in the nonflow 
contributions, which are expected to be larger at high \pt given the 
smaller $\Delta\eta$ gap between the event plane and the track.

\begin{table}
	\caption{\label{tab:respsi2} 
Resolution of $\Psi_2$ measured in the BBCS and FVTXS at each energy and 
centrality.}
	\begin{ruledtabular} \begin{tabular}{cccc}
	\sqsn [GeV] & centrality & $R(\psibbcs)$ & $R(\psifvtxs)$ \\
	\hline
	200  & 0\%--5\%   & 0.1073$\pm$0.0003 & 0.2382$\pm$0.0007 \\
	200  & 5\%--10\%  & 0.085$\pm$0.004  & 0.21$\pm$0.01 \\
	200  & 10\%--20\% & 0.073$\pm$0.003  & 0.168$\pm$0.008 \\
	200  & 20\%--40\% & 0.045$\pm$0.003  & 0.18$\pm$0.01 \\
	200  & 40\%--60\% & 0.031$\pm$0.003  & 0.17$\pm$0.02 \\
	200  & 60\%--88\% & 0.133$\pm$0.003  & 0.22$\pm$0.05 \\
	\\
	62.4  & 0\%--5\%   & 0.0496$\pm$0.0009 & 0.134$\pm$0.002 \\
	62.4  & 5\%--10\%  & 0.0367$\pm$0.0009 & 0.112$\pm$0.003 \\
	62.4  & 10\%--20\% & 0.033$\pm$0.002 & 0.097$\pm$0.006 \\
	62.4  & 20\%--40\% & 0.026$\pm$0.001 & 0.089$\pm$0.004 \\
	62.4  & 40\%--60\% & 0.017$\pm$0.001 & 0.091$\pm$0.006 \\
	62.4  & 60\%--78\% & 0.009$\pm$0.001 & 0.14$\pm$0.02 \\
	\\
	39  & 0\%--10\%  & 0.0255$\pm$0.0009  & 0.069$\pm$0.002 \\
	39  & 10\%--20\% & 0.014$\pm$0.001 & 0.055$\pm$0.005 \\
	39  & 20\%--40\% & 0.010$\pm$0.001 & 0.055$\pm$0.008 \\
	39  & 40\%--60\% & 0.008$\pm$0.002 & 0.037$\pm$0.007 \\
	39  & 60\%--74\% & 0.009$\pm$0.002 & 0.05$\pm$0.01 \\
	\end{tabular} \end{ruledtabular}
\end{table}

At 19.6 GeV, no combination of three-subevents yields a real valued 
event-plane resolution. We expect that this is due to the low 
multiplicity at 19.6 GeV combined with the strong $\eta$ dependence of 
$v_2$. We therefore extrapolate the $R(\psifvtxs)$ from the results at 
higher energies. The event-plane resolution is expected to follow the 
form~\cite{Ollitrault:2009ie}
\begin{equation}
	R(\chi) = \frac{\sqrt{\pi}}{2}\chi e^{-\chi/2}\left[I_0\left(\frac{\chi^2}{2}\right) + I_1\left(\frac{\chi^2}{2}\right)\right],
\end{equation}

where $\chi=v_2\sqrt{N}$, $N$ is the multiplicity, and $I_i$ are the 
modified Bessel functions. The measured resolutions at 200, 62.4, and 39 
GeV are used to extrapolate the resolution at 19.6 GeV under the 
following three assumptions:

\begin{enumerate}

	\item The $v_2$ is constant with \sqsn.

	\item The $v_2$ follows the energy dependence given by the 
\ampt model~\cite{Lin:2004en}, which has been found to reasonably 
reproduce the energy dependence of $v_2$ in small collision 
systems~\cite{Bozek:2015swa,Aidala:2016vgl}.

	\item The $v_2$ follows the energy dependence given by \ampt for 
200--39 GeV, but at 19.6 GeV the $v_2$ is the same as at 39 GeV.

\end{enumerate}

Using the measured multiplicities, we find that all three assumptions 
give results that are in good agreement with the measured resolutions at 
200--39 GeV. We take the average extrapolated resolution from the three 
cases, and assign the maximum extent of the variation as a systematic 
uncertainty. This procedure gives a value of 
$R(\psifvtxs)=0.031^{+0.011}_{-0.016}$ for 0\%--20\% central collisions 
at \sqsn~=~19.6 GeV.

During the \dau data taking in 2016, a 1.0~mrad offset between the 
colliding beams and the longitudinal axis of PHENIX was required due to 
the asymmetric collision species. We negate this effect by applying a 
counter rotation to each central arm track, FVTX cluster, and BBC tube. 
After applying the counter rotation, we find no appreciable offset 
between the $v_2(\pt)$ measured in the east ($\pi/2<\phi<3\pi/2$) and 
west ($-\pi/2<\phi<\pi/2$) central arms for central events. However, as 
we go towards more peripheral events, an increasing difference between 
the east and west central arms is observed. This may be due to a 
decrease in the flow $v_2$ signal relative to background uncorrelated to 
the beam axis. When calculating $\psifvtxs$, we use the standard $Q$ 
vector approach~\cite{Poskanzer:1998yz}. To account for any remaining 
beam offset or background effects, we apply a centrality and collision 
energy dependent offset to the $y$ component of the 2nd order $Q$ 
vector, $\Delta Q_y$, such that the difference between the east and west 
central arms is removed.

The dominant sources of systematic uncertainty in the measurement of 
$v_2(\pt)$ are: (1) Track background from photon conversions and weak 
decays. We estimate the effect of these tracks by comparing the $v_2$ 
measured with a tighter cut on the matching window required for hits in 
the 3rd layer of the PC. We find that this increases the $v_2$ by up to 
2\%, independent of centrality and energy. (2) Contamination from event 
pile-up. The effect of pile-up at 200 GeV is estimated by varying the 
pile-up rejection between $0.92<f<0.98$. This has a negligible effect on 
the $v_2$, and we assign a 1\% uncertainty at 200 GeV. (3) Uncertainty 
on $\Delta Q_y$. As a conservative estimate, we vary the $\Delta Q_y$ 
values by $\pm50\%$ and compare the resulting $v_2(\pt)$ values. An 
uncertainty of $<1\%$--$9\%$ that varies with energy and centrality is 
assigned based on the study. (4) The difference between the $v_2(\pt)$ 
values measured independently using the FVTXS and BBCS event planes. As 
discussed above, this difference for $\pt<2$ GeV/$c$ is found to be 
2.5\% independent of centrality and energy. (5) The difference between 
the event-plane and two-particle-correlation methods. As shown in 
Fig.~\ref{fig:v2meth}, there is good agreement between the two methods 
in central collisions, however there is some difference for more 
peripheral collisions. We include this difference as an additional 
systematic uncertainty. (6) Uncertainty in the event-plane resolution as 
given in Table~\ref{tab:respsi2}. As discussed above, the resolution at 
19.6 GeV is extrapolated from the measured results at 200--39 GeV and a 
systematic uncertainty is assigned based on varying the assumptions of 
the extrapolation. The uncertainties are summarized in 
Table~\ref{tab:syspt}, categorized by $type$. PHENIX considers three 
categories of systematic uncertainties:

\begin{enumerate}
	\item \textbf{Type A}: point-to-point uncorrelated;
	\item \textbf{Type B}: point-to-point correlated;
	\item \textbf{Type C}: global scale uncertainties.
\end{enumerate}

On all plots, type A uncertainties are represented as vertical error 
bars, type B uncertainties by filled boxes, and type C uncertainties are 
quoted on the plot or in the legend.

\begin{table*}
	\caption{\label{tab:syspt} 
Systematic uncertainties on measurements of $v_2$ vs \pt.}
	\begin{footnotesize}
\begin{ruledtabular} \begin{tabular}{ccrrrr}
	\multirow{2}{*}{Source} & \multirow{2}{*}{Type} & \multicolumn{4}{c}{\sqsn [GeV]} \\
	       &      & 200 & 62.4 & 39 & 19.6 \\
	\hline
	Track Background       & B & 2.0\% & 2.0\% & 2.0\% & 2.0\% \\
	Event Pile-up          & B & 1.0\% & $<1\%$ & $<1\%$ & $<1\%$ \\
	Beam Angle             & B & $<1\%$--5\% & $<1\%$--9\% & $<1\%$--8\% & $<1\%$ \\
	Event-Plane Detector   & B & 2.5\% & 2.5\% & 2.5\% & 2.5\% \\
	Event-Plane Method     & B & 0.4\%--17.5\% & 0.4\%--17.5\% & 1.6\%--17.5\% & 6.2\% \\
	Event-Plane Resolution & C & 0.3\%--23.0\% & 1.8\%--12.8\% & 3.6\%--20.4\% & $^{+35\%}_{-48\%}$ \\
	\end{tabular} \end{ruledtabular}
	\end{footnotesize}
\end{table*}

In previous PHENIX publications on flow in small 
systems~\cite{Aidala:2016vgl,Adare:2015ctn}, an estimation of the 
nonflow contributions to the measured $v_2$ has been included in the 
systematic uncertainties. The estimation used the ratio of the $C_2$ 
measured in \pp collisions, scaled by the relative charge in the BBCS, 
to the $C_2$ measured in \pdheau. In Ref.~\cite{Adare:2014keg}, nonflow 
was estimated to contribute positively between $\sim5\%$ at $\pt=1$ 
GeV/$c$ and $\sim10\%$ at $\pt=4$ GeV/$c$ to the observed $v_2$ signal. 
This estimation assumes that correlations in \pp collisions come from 
nonflow alone, which may be an overestimate given recent results in \pp 
collisions at the LHC. In this analysis we lack a suitable \pp reference 
at all four energies and, therefore, do not make any estimation of the 
nonflow contributions to the measured $v_2$ in this paper.

Figure~\ref{fig:v2meth} shows the $v_2$ vs \pt in 0\%--5\% central \dau 
collisions at \sqsn~=~200 GeV measured with the event-plane method 
compared to the two-particle correlation method described above.  The 
two methods are consistent with each other. The two-particle method 
always gives the RMS average of $v_2$, i.e. $\sqrt{\langle 
v_2^2\rangle}$.  By contrast, the event-plane method is an estimator of
$\langle v_2^{\alpha}\rangle^{1/\alpha}$~\cite{Ollitrault:2009ie}, 
where $1<\alpha<2$.  For sufficiently high-multiplicities, e.g. in 
central $A$$+$$A$, $\alpha$ approaches 1 and the event-plane method is 
an estimator of $\langle v_2\rangle$. As the multiplicity decreases, 
$\alpha$ approaches 2 and the event-plane method is equivalent to the 
two-particle method.  The consistency between the two methods here 
demonstrates we are in the regime where the multiplicity is low enough 
that the two methods are equivalent.  It is important to remember, 
then, that all event-plane method results have the same dependence on 
the fluctuations of the $v_2$ distribution as the 2-particle method.

Also shown in Fig.~\ref{fig:v2meth} is the previously published 
measurement of $v_2(\pt)$ in 0\%--5\% central \dau collisions at 
\sqsn~=~200 from PHENIX using data collected in 
2008~\cite{Adare:2014keg}. The results are in good agreement for $\pt<2$ 
GeV/$c$. We note that the result presented here uses a different 
detector to measure the event plane than that used in 
Ref.~\cite{Adare:2014keg}. This is a dominant source of systematic 
uncertainty in the measurement and is therefore largely uncorrelated 
between the two. Further, at high \pt, nonflow effects play a larger 
role (as discussed later in this paper), and are dependent on the 
$\Delta\eta$ gap between the region in which the event plane is measured 
and the region in which the $v_2$ is measured. The increasing nonflow at 
high \pt, which is not estimated in the measurement presented here, 
potentially explains the modest difference between the two measurements.

\subsection{Analysis of $v_2$ vs $\eta$ using the event-plane method}
\label{sec:v2eta}

The measurement of the $\eta$ dependence of $v_2$ uses the same 
event-plane method as discussed in Sec.~\ref{sec:v2pt}. However, in 
order to cover the maximum extent in $\eta$, tracks in both the FVTXN 
and FVTXS are included alongside tracks measured in the central arms. 
This necessitates using the event plane measured in the BBCS 
($\psibbcs$), rather than the FVTXS. The resolutions of $\psibbcs$ at 
each energy are given in Table~\ref{tab:respsi2}.

\begin{figure}[htbp]
	\includegraphics[width=1.0\linewidth]{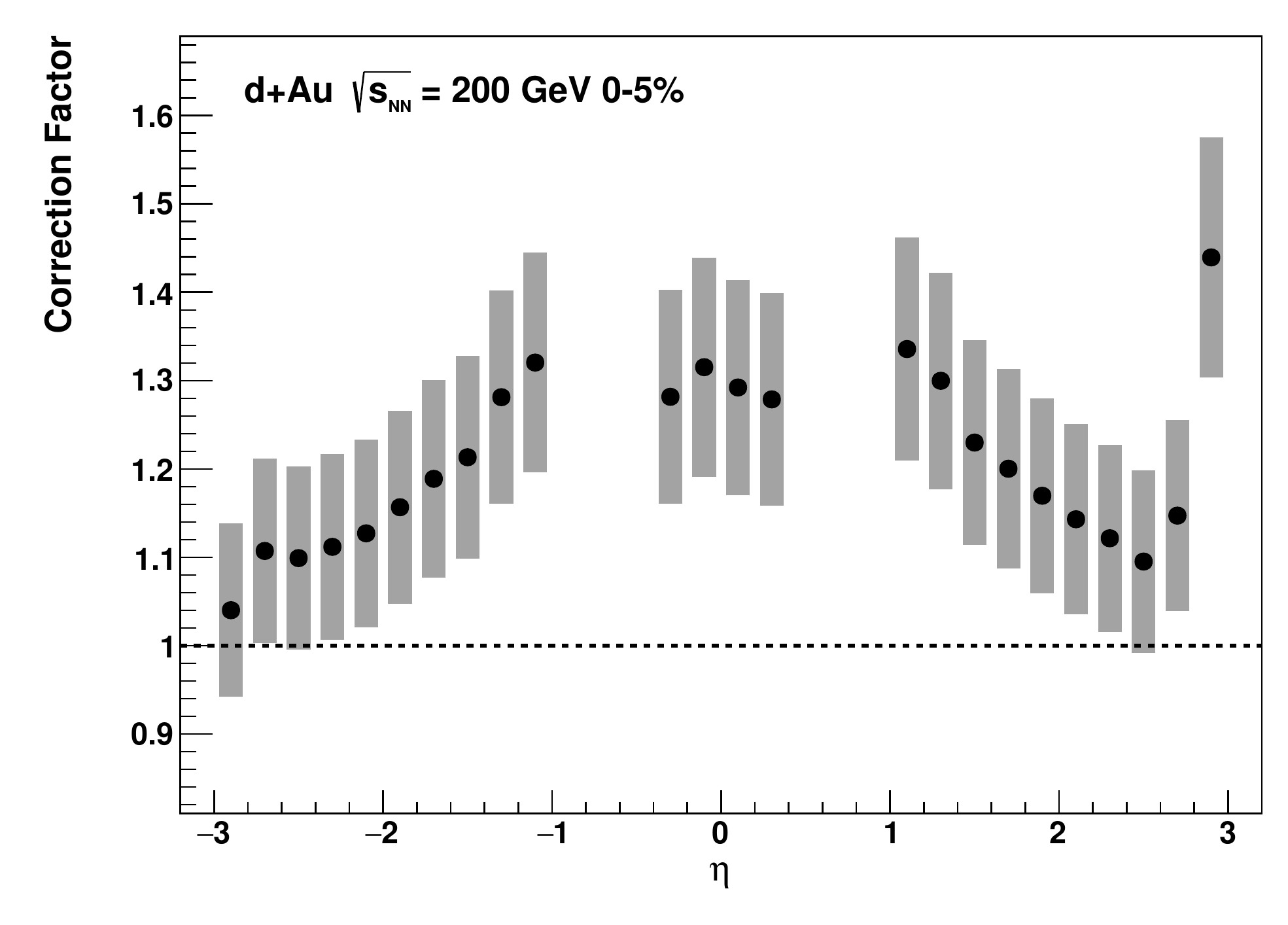}
	\caption{\label{fig:v2etacorr} 
The correction factor on $v_2(\eta)$ as a function of $\eta$.}
\end{figure}

To calculate the \pt-integrated $v_2(\eta)$ we must correct for 
the detector acceptance and efficiency. This correction is estimated 
using the (\ampt) model~\cite{Lin:2004en}, coupled to 
a full \geant model~\cite{GEANT} of the PHENIX detector. The ``true'' 
$v_2$ is calculated in \ampt relative to the parton participant plane, 
$\psipp$. The same events are then run through \geant and the $v_2$ is 
recalculated using reconstructed tracks, relative to the same $\psipp$. 
The resulting correction factor ($\varepsilon^{\rm corr}(\eta)$) for 
\dau at \sqsn~=~200 GeV is shown in Fig.~\ref{fig:v2etacorr}, and is 
found to range from 2\%--30\%. The correction factors at 62.4 and 39 GeV 
are similar, but show systematic increases at forward rapidity. The 
uncertainty on the correction factor is estimated by investigating the 
following effects:

\begin{enumerate}
	\item The correction's dependence on the true $v_2$.
	\item The correction's dependence on the true \pt distribution.
	\item The correction's dependence on the simulation-to-data matching.
\end{enumerate}

We investigate the correction's dependence on the true $v_2$ by varying 
the parton-parton interaction cross section in \ampt from 1.5 mb to 3.0 
mb. This causes a change in the true $v_2$ of $\sim20\%$. The correction 
factor is found to change by a maximum of $\pm3\%$. To test the 
correction factor's sensitivity to the true \pt distribution, the shape 
of the input \pt distribution is modified such that the mean \pt changes 
by $\pm20\%$. We find that this changes the correction factor by 
$\pm8\%$. Finally, we test the correction's sensitivity to the detailed 
detector acceptance and efficiency by making tight fiducial cuts, 
including only regions that agree well between data and simulations. 
This leads to a maximum change in the correction factor of $\pm4\%$. 
Adding these in quadrature, a $\pm9.4\%$ systematic uncertainty is 
assigned on the correction factor. This leads to a systematic 
uncertainty on the measured $v_2(\eta)$ of $<1$--3\% that varies with 
$\eta$.

The resulting, \pt-integrated, $v_2(\eta)$ is calculated using
\begin{equation}
	v_2(\eta) = \frac{\langle\cos{2(\phi_{\rm trk}(\eta) - \psibbcs)}\rangle}{R(\psibbcs)\varepsilon^{\rm corr}(\eta)},
	\label{eq:v2eta}
\end{equation}
where $\phi_{\rm trk}$ is the azimuthal angle of tracks in the FVTX or 
central arms, $\psibbcs$ is the second-order azimuthal event plane 
measured by the BBCS, $R(\psibbcs)$ is the resolution of $\psibbcs$, and 
$\varepsilon^{\rm corr}(\eta)$ is the detector acceptance and 
efficiency correction factor.

The other dominant sources of systematic uncertainty are similar to 
those detailed for the $v_2(\pt)$ measurement above. (1) Track 
background in the FVTX is investigated by tightening the $DCA$ track 
cut. We assign a 2\% uncertainty on $v_2(\eta)$ based on this study. (2) 
The same 1\% systematic uncertainty due to event pile-up is assigned 
based upon the investigation detailed in Sec.~\ref{sec:v2pt}. (3) 
Remaining effects due to the 1.0 mrad beam angle are investigated by 
looking at the difference in the $v_2(\eta)$ as measured by the east and 
west central arms. We estimate a systematic uncertainty on $v_2(\eta)$ 
assuming a uniform distribution as $\sigma=\sqrt{\langle 
v_2^{\rm west}(\eta)\rangle - \langle 
v_2^{\rm east}(\eta)\rangle}/\sqrt{12}$, which is found to vary with 
collision energy between 6.5\%--33.9\%. (4) As in the measurement of 
$v_2(\pt)$, we cross check the result, which in this case uses the BBCS 
event plane, with $v_2(\eta)$ measured using the FVTXS event plane. This 
allows us to test the agreement at mid and forward rapidities, but not 
at backward rapidity because tracks cannot be measured in the same region 
in which the event plane is measured. We find a larger difference 
between the event-plane results in the forward region and assign a 6.5\% 
uncertainty based on the difference. (5) A systematic uncertainty is 
assigned based on the uncertainty in the calculated event-plane 
resolution, as given in Table~\ref{tab:respsi2}. A summary of the 
systematic uncertainties, and their assigned type, is shown in 
Table~\ref{tab:syseta}.

\begin{table}
	\caption{\label{tab:syseta} 
Systematic uncertainties on measurements of $v_2$ vs $\eta$.}
\begin{ruledtabular} \begin{tabular}{ccrrr}
	\multirow{2}{*}{Source} & \multirow{2}{*}{Type} & \multicolumn{3}{c}{\sqsn [GeV]} \\
	& & 200 & 62.4 & 39 \\
	\hline
	Track Background       & B & 2.0\%  & 2.0\%  & 2.0\%  \\
	Event Pile-up          & B & 1.0\%  & $<1$\% & $<1$\% \\
	east vs west           & B & 4.3\%  & 13.4\% & 33.9\% \\
	Event-Plane Detector   & B & 6.5\%  & 6.5\%  & 6.5\%  \\
	Efficiency correction  & B & 0--3\% & 0--3\% & 0--3\% \\
	Event-Plane Resolution & C & 0.3\%  & 1.8\%  & 3.6\%  \\
	\end{tabular} \end{ruledtabular}
\end{table}

\subsection{Analysis of $dN_{ch}/d\eta$ vs $\eta$}
\label{sec:dndeta}

We begin by measuring the ratio of \dndeta in central \dau collisions at 
\sqsn~=~62.4, 39, and 19.6 GeV relative to 0\%--5\% central \dau 
collisions at \sqsn~=~200 GeV. The ratio of the raw track distributions 
are calculated using the analysis cuts described in 
Sec.~\ref{sec:experiment}. Variations in the detector performance over 
time, and as a function of the azimuthal angle, are tested by selecting 
ten different time periods during the data taking at each energy, as 
well as four distinct regions in $\phi$. The RMS of the ratios for each 
combination of time period and $\phi$ range are taken as a systematic 
uncertainty.

\begin{figure}[htbp]
	\includegraphics[width=1.0\linewidth]{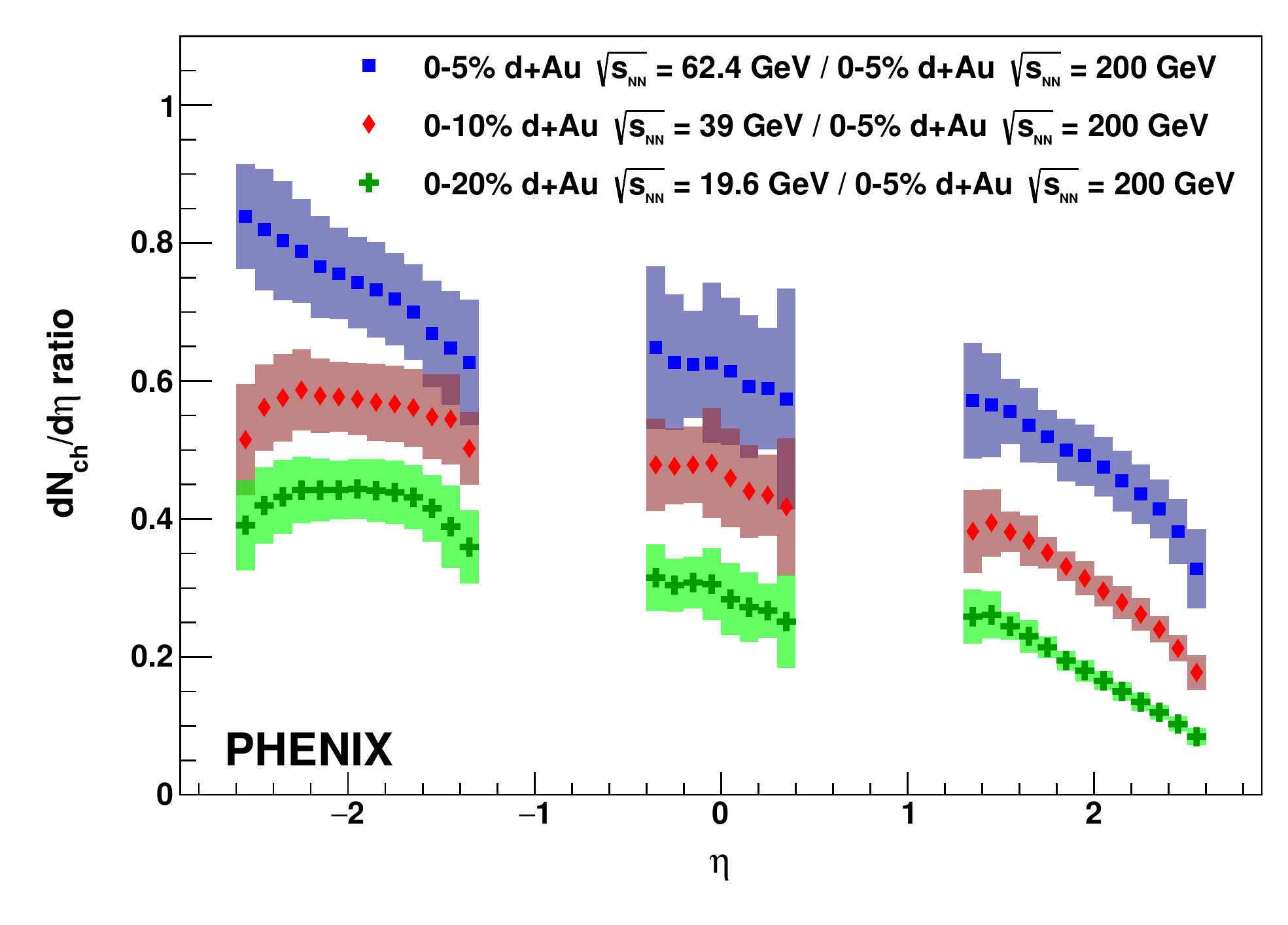}
	\caption{\label{fig:dndeta_ratio} 
Ratio of \dndeta vs $\eta$ in central \dau collisions at \sqsn~=~62.4, 
39, and 19.6 GeV relative to 0\%--5\% central \dau collisions at 
\sqsn~=~200 GeV. Systematic uncertainties are shown as filled boxes 
surrounding each point.}
\end{figure}

\begin{figure*}[htbp]
    \includegraphics[width=0.99\textwidth]{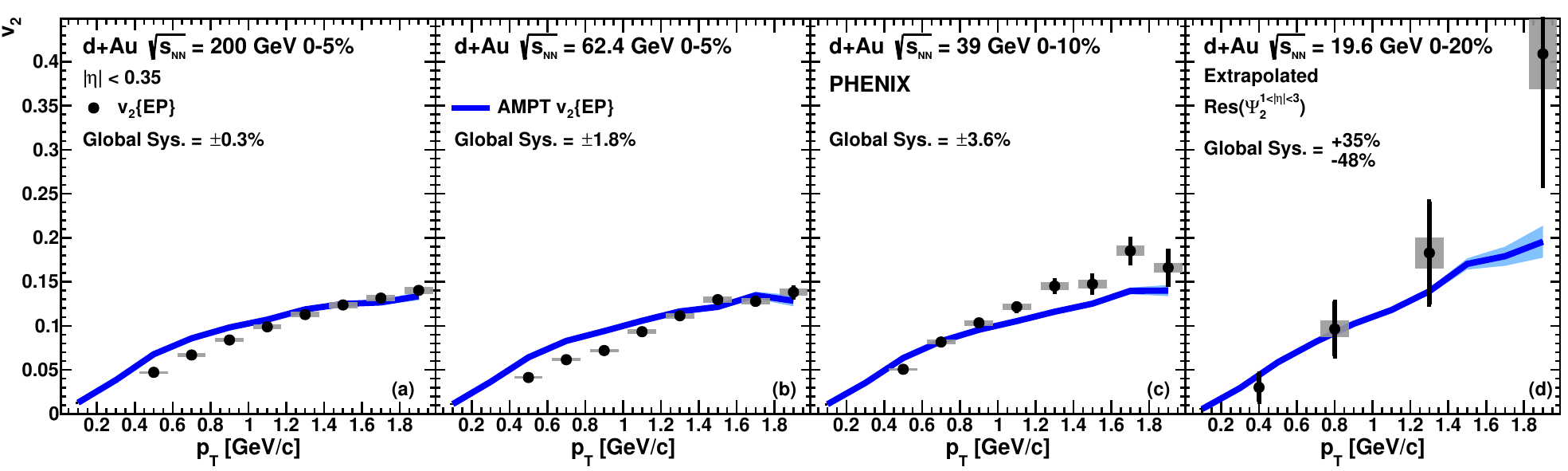}
    \includegraphics[width=0.99\textwidth]{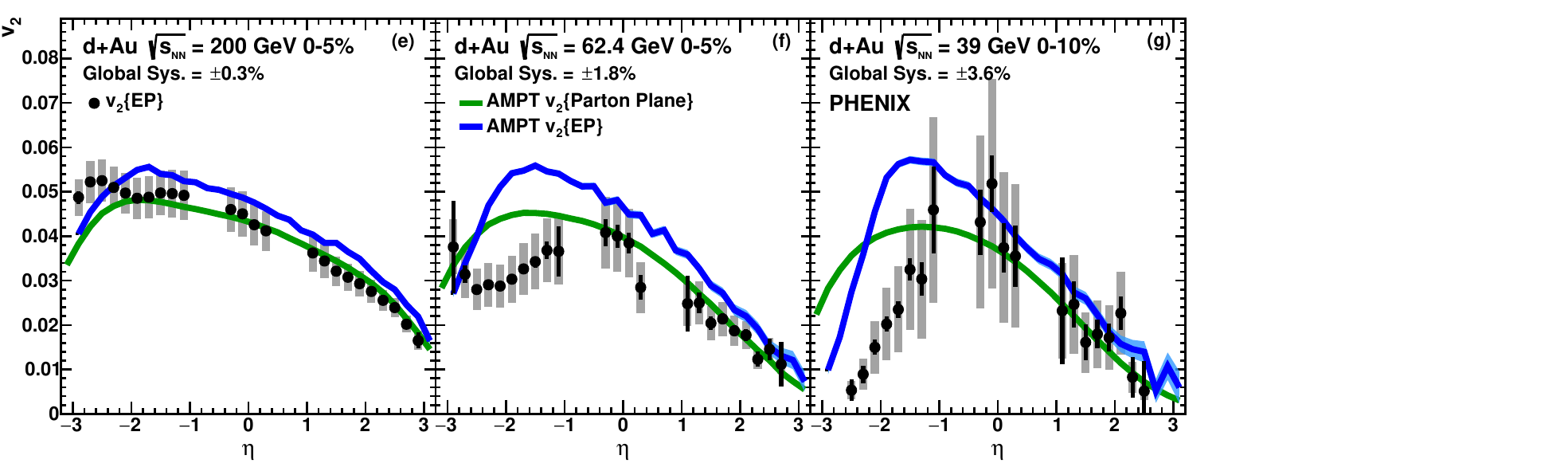}
    \caption{\label{fig:v2_ampt} 
The value of $v_2$ as a function of \pt in central \dau collisions at 
\sqsn~=~200 (a), 62.4 (b), 39 (c), and 19.6 (d) GeV. $v_2$ as a function 
of $\eta$ in central \dau collisions at \sqsn~=~200 (e), 62.4 (f), and 
39 (g) GeV. The lower [green] curves show calculations from the \ampt 
model~\cite{Lin:2004en} where the $v_2$ is calculated relative to the 
parton plane. The upper [blue] curves show calculations from the \ampt 
model, where the $v_2$ is calculated using the event-plane method, 
as described in the text.}
\end{figure*}

When calculating the \dndeta ratios, it is also important to consider 
the change in acceptance and efficiency ($A\times\varepsilon$) between 
collision energies, particularly due to changes in the mean \pt 
(\meanpt). We calculate the change in $A\times\varepsilon$ by simulating 
\ampt events at each collision energy, run through a full \geant 
description of the PHENIX detector. The ratio of the resulting 
$A\times\varepsilon$ distributions for each energy are then calculated 
as a correction to the ratios in raw data. The sensitivity of the 
$A\times\varepsilon$ ratio to the true \pt distribution is tested by 
varying the relative \meanpt between energies by $\pm$10\%. This yields 
a maximum change in the $A\times\varepsilon$ ratio of 10\%, which we 
assign as a systematic uncertainty. The corrected \dndeta ratios are 
shown in Fig.~\ref{fig:dndeta_ratio}.

To calculate the absolutely normalized \dndeta at each energy, 
we fix the \dndeta in 0\%--20\% \dau collisions at \sqsn~=~200 GeV to 
the result previously measured by PHOBOS~\cite{Back:2004mr}. The PHOBOS 
result is in excellent agreement with the previously published \dndeta 
at midrapidity measured by PHENIX~\cite{Adare:2015bua}. This method 
allows us to reduce the overall systematic uncertainties that arise from 
calculating an absolutely normalized $A\times\varepsilon$. To calculate 
the \dndeta in 0\%--5\% central \dau collisions at \sqsn~=~200 GeV, we 
also need the ratio of \dndeta in 0\%--5\% / 0\%--20\% central \dau 
collisions at 200 GeV. This ratio is calculated in the same manner 
described above. The systematic uncertainties on the PHOBOS measurement 
are propagated directly to the \dndeta in 0\%--5\% central \dau 
collisions at 200 GeV.

\section{Results and Discussion}
\label{sec:discussion}

The $v_2(\pt)$ in central \dau collisions at \sqsn~=~200, 62.4, 39, and 
19.6 GeV is shown in Fig.~\ref{fig:v2_ampt}(a)--(d). The $v_2(\pt)$ in 
centrality bins are shown in Appendix~\ref{appendix:A}. A positive $v_2$ 
signal that increases with increasing \pt is observed in all centrality 
bins at all four energies.

The $v_2(\eta)$ in central \dau collisions at \sqsn~=~200, 62.4, and 39 
GeV is shown in Fig.~\ref{fig:v2_ampt}(e)--(g). At all three energies we 
observe a $v_2$ that decreases with increasing $\eta$ between 
$0<\eta<3$. At 200 GeV, the $v_2$ at backward rapidity is similar or 
greater to that measured at $\eta=0$. This is reminiscent of the 
asymmetric $dN_{ch}/d\eta$ measured in \dau 
collisions~\cite{Back:2004mr}. At 62 GeV the $v_2$ at backward rapidity 
starts to decrease for $\eta<0$. This trend is stronger at 39 GeV, where 
the $v_2$ distribution falls to zero for $\eta=-2.8$. This decrease at 
backward rapidity may be due to nonflow contributions in regions near 
where the event plane is measured ($-3.9<\eta<-3.1$ in this case). This 
possibility is discussed in more detail in Sec.~\ref{sec:ampt}.

\subsection{Comparison of $v_2$ results with \ampt calculations}
\label{sec:ampt}

The (\ampt) model~\cite{Lin:2004en} combines string melting and then both 
partonic and hadronic scattering.  It has previously been compared to 
measurements of flow in small collision 
systems~\cite{Adare:2015ctn,Aidala:2016vgl,Bozek:2015swa,Koop:2015wea}, 
and found to be in good agreement with $p/d/^3$He$+$Au collisions at 
\sqsn~=~200 GeV for $\pt<1$ GeV/$c$. Following Ref.~\cite{Koop:2015trj}, 
we use \ampt Version 2.26, which is additionally modified to utilize the 
Hulth\'en wavefunction description of the deuteron and black disk 
nucleon-nucleon interactions with the Monte-Carlo Glauber component. 
Further details are discussed in Appendix~\ref{appendix:B}. In addition, 
within \ampt one can run with only partonic scattering (i.e. no hadronic 
scattering) or with only hadronic scattering (i.e. no partonic 
scattering), and the results are also shown in 
Appendix~\ref{appendix:B}. In all cases, the charged particle 
multiplicity in the region $-3.9<\eta<-3.1$ is used to determine the 
event centrality class in a manner consistent with the experimental 
measurements. We begin the discussion by focusing on the most central 
collisions, as shown in Fig.~\ref{fig:v2_ampt}, and return to the full 
centrality dependence later.

\subsubsection{Central collisions}

Figure~\ref{fig:v2_ampt} shows the $v_2$ calculated relative to the 
$\Psi_2$ plane calculated from initial partons, labeled 
$\vtpp$~\footnote{We note that Ref.~\cite{Koop:2015trj} includes \ampt 
calculations of $v_2(\pt)$ relative to the initial \textit{nucleon} 
positions for $b<2$ fm \dau collisions at the energies measured here. 
The results are broadly similar to those shown here.}. By calculating 
$v_2$ relative to the parton plane, we can isolate the $v_2$ that is 
truly coupled to the initial geometry, or what we refer to as flow. At 
200 and 62.4 GeV, \ampt provides a reasonable description of the data 
for $\pt<1$ GeV/$c$ and under-predicts the data for $\pt>1$ GeV/$c$. At 
39 and 19.6 GeV \ampt under-predicts the data at all but the lowest \pt. 
We further find good agreement between $v_2(\eta)$ and $\vtpp$ at mid 
and forward rapidities at all three collision energies. At backward 
rapidity we find good agreement at 200 GeV, but \ampt does not show the 
same fall-off as seen in the data at 62.4 and 39 GeV.

Because \ampt is a full event generator, we can not only determine 
$\vtpp$, but also mimic in detail the experimental measurement using 
only the final-state particles. We use the same event-plane method as 
used in the data analysis, matching the nominal pseudorapidity ranges of 
the detectors rather than a full \geant simulation of the detector 
response. This result, labeled as $\vtep$, includes not only flow, but 
also nonflow correlations as modeled within \ampt. The results are shown 
in Fig.~\ref{fig:v2_ampt}. As a function of \pt, the $\vtep$ 
calculations are similar to $\vtpp$ for $\pt<0.5$ GeV/$c$. For $\pt>0.5$ 
GeV/$c$ the event-plane results produce a larger $v_2$ signal, which is 
in better agreement with the data. This difference highlights the 
contributions from nonflow that, in \ampt, increase with increasing \pt 
and decreasing collision energy.

\begin{figure*}[bth]
	\includegraphics[width=0.99\textwidth]{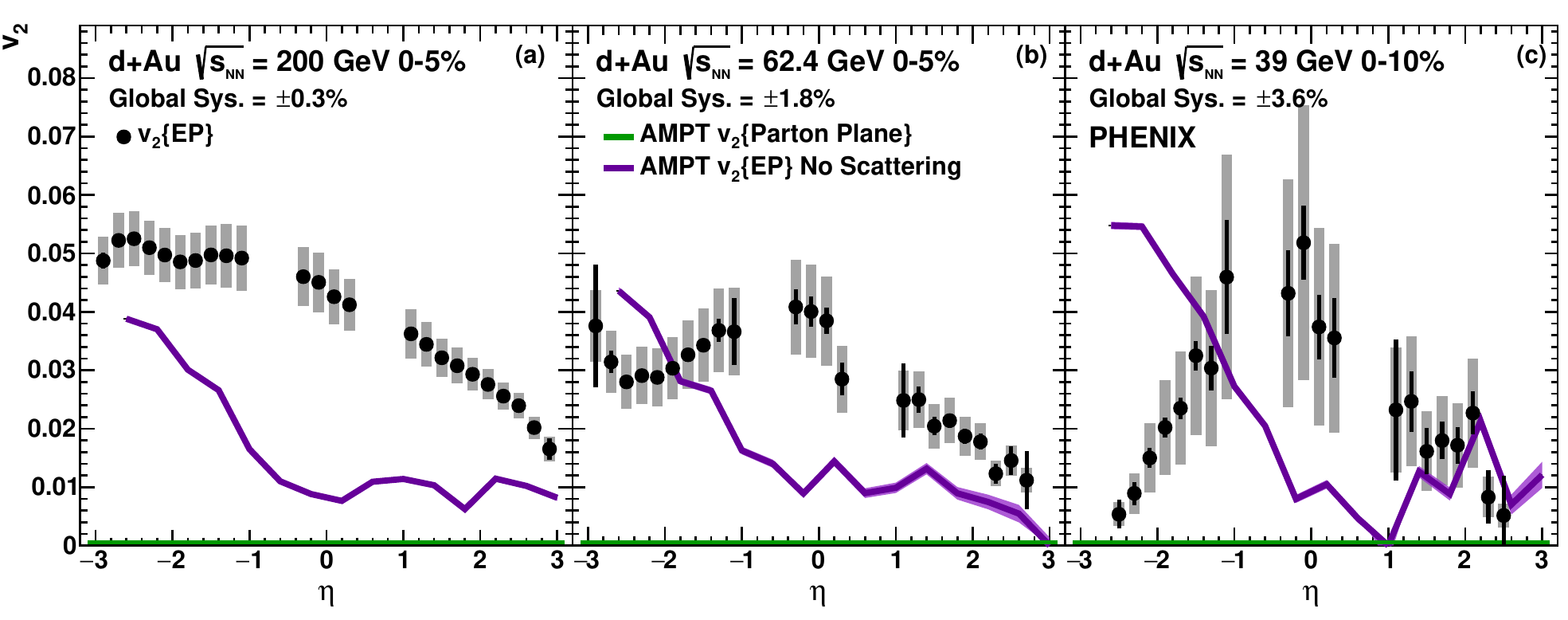}
	\caption{\label{fig:v2etanint} 
The value of $v_2$ as a function of $\eta$ in central \dau collisions at 
\sqsn~=~200, 62.4, and 39 GeV compared to calculations from the \ampt 
model~\cite{Lin:2004en} in which both the partonic and hadronic 
scatterings have been turned off. The upper [purple] curves show 
calculations from the \ampt model using the event-plane method as 
described in the text.  The lower [green] curves ($v_{2}=0$ in all cases)
show calculations from \ampt where $v_2$ is calculated relative 
to the parton plane.
}
\end{figure*}

When looking at $v_2(\eta)$, shown in Fig.~\ref{fig:v2_ampt}, we find 
that the \ampt event-plane results are in good agreement with the 
measured data for $\eta>0$ at all three collision energies. At 200 GeV 
we see a roughly constant increase vs $\eta$ of the $\vtep$ compared to 
the $\vtpp$, indicating a roughly $15\%$ increase in the $v_2$ with the 
addition of nonflow. Both calculations are in agreement with the data 
within uncertainties. At 62.4 and 39 GeV we see a larger increase in the 
event-plane result versus the parton plane result compared to the 200 
GeV. What is particularly interesting is that \ampt shows a decrease in 
the event-plane result for $\eta<-2$ that is stronger for 39 GeV than 
62.4 GeV, and drops below the parton plane result at $\eta\approx-2.5$. 
While this decrease doesn't occur at the same $\eta$, and is only in 
qualitative agreement with the data, it points out that within \ampt 
this feature only arises when you combine flow and nonflow. When using 
the event-plane method at these low energies, \ampt predicts a larger 
deviation between the true flow signal and the experimentally observed 
flow signal as the $\Delta\eta$ between the region in which the tracks 
are measured and the region in which the event plane is measured 
decreases. We further caution that, while \ampt qualitatively agrees 
with our measurements over a broad range in collision energy and 
particle kinematics, we can not use it to definitively separate flow 
from nonflow, but rather to give some insight and possible intuition for 
interpreting the experimental results in regions where we are currently 
unable to perform the separation experimentally.

Using \ampt, we can also study whether our measured $v_2(\eta)$ is 
likely to arise solely from nonflow contributions. By setting the 
partonic and hadronic interaction cross sections to zero within \ampt, 
we eliminate all interactions that translate initial-state geometry to 
final-state momentum correlations. This is shown explicitly in 
Fig.~\ref{fig:v2etanint}, where $\vtpp=0$ at all $\eta$. However, even 
with all partonic and hadronic scattering turned off, nonflow 
correlations can still give rise to a $\vtep$ signal. This is shown by 
the upper [purple] curves in Fig.~\ref{fig:v2etanint}. Note, that in this mode 
the event plane angle arises only from nonflow correlations, and has no 
connection to the initial geometry (i.e. the parton plane). In this case 
the resolution of the event plane is roughly a factor of 3 lower than 
with partonic and hadronic interactions. At all three energies, 
$\vtep<0.01$ for $\eta>0$ within \ampt with partonic and hadronic 
scattering switched off. This region is far removed ($\Delta\eta>3.1$) 
from the region in which the event plane is constructed and is therefore 
unlikely to contain correlations from jets or particle decays. In the 
region $\eta<0$, however, an increasing $\vtep$ is observed. This 
indicates, as expected, that the smaller the $\Delta\eta$ gap the larger 
the effects of nonflow. In all cases, the measured $v_2(\eta)$ for 
$\eta>0$ is larger than the $\vtep$ from \ampt with nonflow correlations 
only. This extends to $\eta<0$ in central collisions at 200 GeV. The 
small values of the $\vtep$ from nonflow correlations only lends further 
confidence that the low \pt and $\eta>0$ region is dominated by flow 
correlations linked to the initial geometry of the collision. We note 
that it is not clear how this large increasing $\vtep$ signal at 
$\eta<0$ with partonic and hadronic scattering turned off (nonflow only) 
turns into a decreasing $\vtep$ signal at $\eta<0$ when partonic and 
hadronic scattering are turned on (flow $+$ nonflow). Presumably this is 
due to detailed interactions between the angle of the parton plane and 
the dominant axis of the nonflow on an event-by-event level within 
\ampt.

\subsubsection{Centrality dependence}

\begin{figure*}[htbp]
	\includegraphics[width=0.75\linewidth]{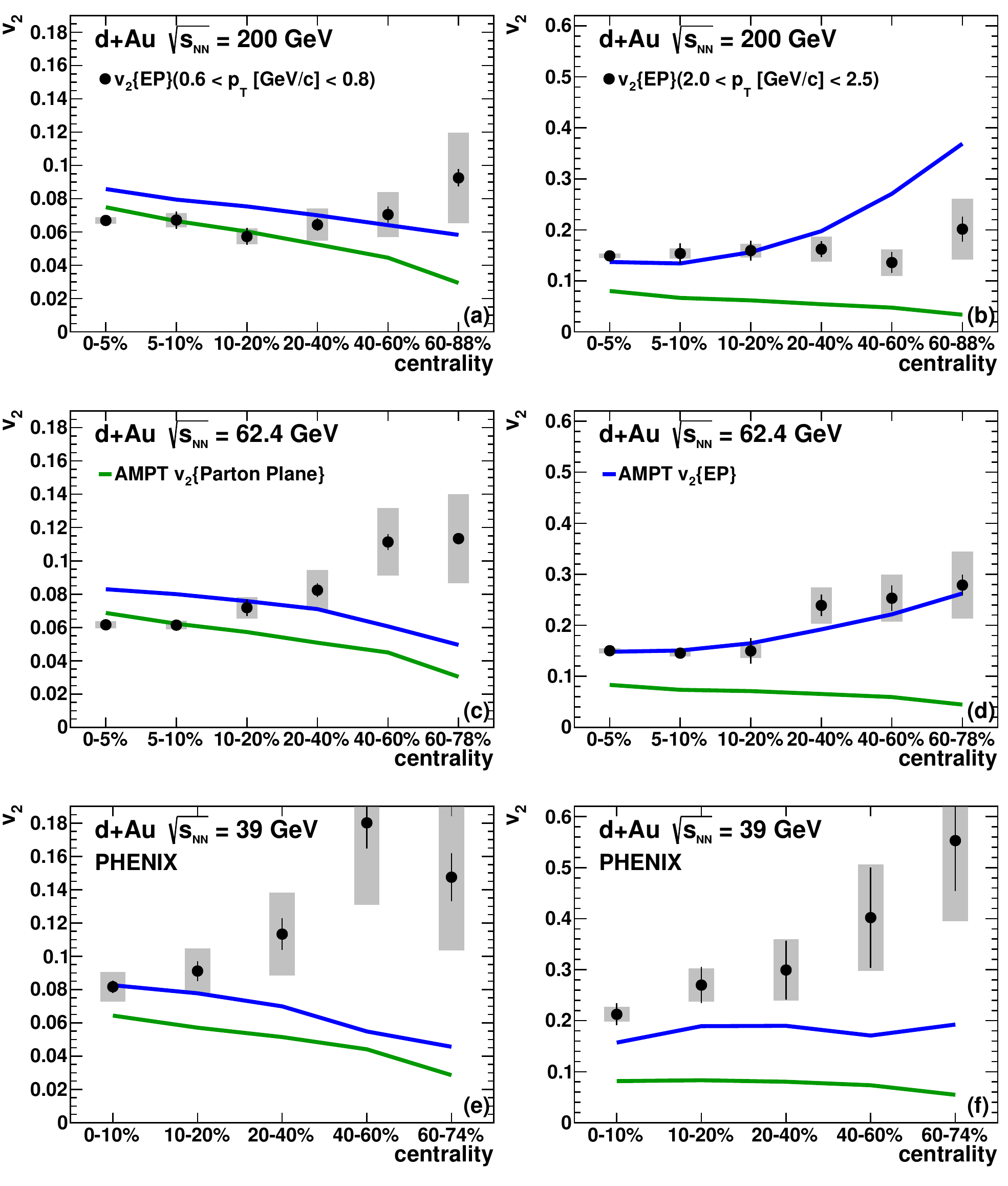}
	\caption{\label{fig:v2cent} 
The value of $v_2$ as a function of centrality  
(a,c,e) at $0.6<\pt[{\rm GeV}/c]<0.8$ and 
(b,d,f) at $2.0<\pt\ [{\rm GeV}/c]<2.5$ in \dau collisions 
at \sqsn~=~(a,b) 200, (c,d) 62.4, and (e,f) 39 GeV. The upper [blue] 
curves show calculations from the \ampt model~\cite{Lin:2004en} where 
the $v_2$ is calculated using the event plane method as described in the 
text.  The lower [green] curves show \ampt calculations where the $v_2$ 
is calculated relative to the parton plane.
}

\end{figure*}

We now return to the centrality dependence of $v_2(\pt)$. From the 
comparison of $v_2(\pt)$ in central collisions we can separate the \pt 
spectra into two regions: (1) $\pt<1$ GeV/$c$ where \ampt parton and 
event plane results are roughly similar. (2) $\pt>1$ GeV/$c$ where the 
event plane results, which include nonflow contributions, yield a larger 
$v_2$ than that calculated with the parton plane. We choose two 
particular \pt bins, $0.6<\pt<0.8$ and $2.0<\pt<2.5$, and investigate 
the centrality dependence of the $v_2$ at \sqsn~=~200, 62.4, and 39 GeV 
in comparison with the results from \ampt, as shown in 
Fig.~\ref{fig:v2cent}. Note that while the event plane resolution 
uncertainty is a global scale uncertainty when plotting $v_2$ as a 
function of \pt, when plotting $v_2$ as a function of centrality it 
becomes a type B systematic uncertainty and is added in quadrature with 
the other type B systematic uncertainties in Fig.~\ref{fig:v2cent}.

Starting with the low \pt $v_2$, \ampt shows similar results between the 
parton and event planes, indicating within \ampt that the flow dominates 
in this \pt region. The \ampt results also predict a decrease in the 
$v_2$ results towards more peripheral collisions, as expected from the 
decrease in the mean ellipticity of the initial geometry and lower 
particle multiplicity. This is contrary to the trends in the data where 
the values of $v_2$ increase in the most peripheral collisions. This 
increase is more pronounced in the lower-energy data and it may indicate 
that nonflow contributions are larger in the data than in AMPT. The 
$v_2$ values measured in the centrality range up to 20\% are in good 
agreement with the predictions from AMPT.

At high \pt, \ampt predicts a significantly larger $v_2$ calculated 
relative to the event plane compared to the parton plane, indicating 
significant contributions from nonflow correlations. At 39 and 62.4 GeV, 
we observe a $v_2$ that increases with more peripheral collisions. At 
62.4 GeV, \ampt well reproduces this increasing behavior. At 200 GeV, 
however \ampt over-predicts the observed increase, while 
under-predicting the increase at 39 GeV.

\subsection{Comparison of $v_2$ results with hydrodynamic calculations}
\label{sec:hydro}

\begin{figure*}[htbp]
    \includegraphics[width=0.99\textwidth]{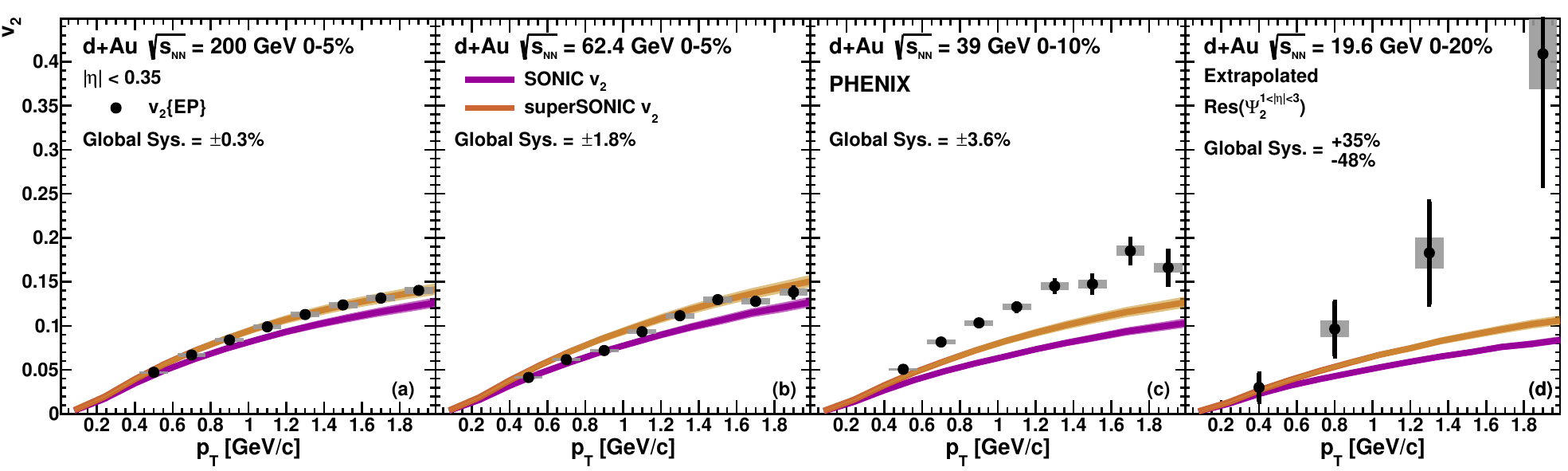}
    \includegraphics[width=0.99\textwidth]{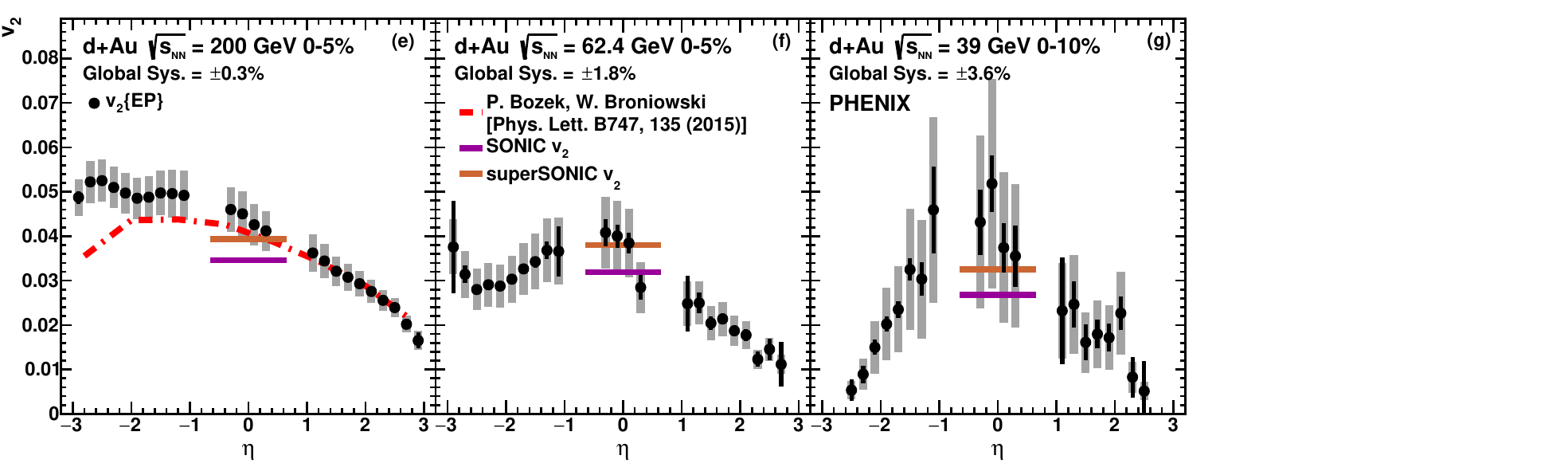}
    \caption{\label{fig:v2_theory} 
The value of $v_2$ as a function of \pt in central \dau collisions at 
\sqsn~=~(a) 200, (b) 62.4, (c) 39, and (d)~19.6~GeV. $v_2$ as a function 
of $\eta$ in central \dau collisions at \sqsn~=~(e) 200, (f) 62.4, and 
(g)~39~GeV.  At midrapidity, the [lower] purple and upper [orange] 
curves show theoretical calculations from \sonic and 
\supersonic~\protect\cite{Koop:2015trj}, respectively.  The dashed [red] 
curve in panel (e) shows hydrodynamic predictions 
from Ref.~\protect\cite{Bozek:2014cya}.}
\end{figure*}

Shown in Fig.~\ref{fig:v2_theory} are predictions from the \sonic and 
\supersonic models for $v_2(\pt)$ at midrapidity~\cite{Koop:2015trj}. 
The \sonic model~\cite{Habich:2014jna} uses Monte-Carlo Glauber initial 
conditions to determine the energy density distribution.  For these 
calculations, $b<2$ fm was used to represent the central-event category. 
While $b<2$ fm is not a direct match for our central multiplicity bins, 
the resulting $\varepsilon_2$ values are consistent with those given in 
Table~\ref{tab:cent}. The initial energy density is tuned such that the 
$dN_{ch}/d\eta$ at midrapidity matches the values given in 
Table~\ref{tab:dndeta}. The Glauber initial conditions are followed by 
viscous hydrodynamics with $\eta/s=1/4\pi$, and at $T=170$ MeV the 
transition to a hadron cascade. The \supersonic 
model~\cite{Romatschke:2015gxa} additionally includes pre-equilibrium 
dynamics. At 200 and 62.4 GeV, both calculations are in excellent 
agreement with the data, with \supersonic providing a slightly better 
description for $\pt>1$ GeV/$c$. At 39 and 19.6 GeV, both calculations 
under-predict the data for $\pt>0.5$ GeV/$c$. This difference may be due 
to the increasing contributions of nonflow present in the data at high 
\pt and lower collision energies, which is not accounted for in these 
calculations. Without a reliable estimate of the nonflow contribution, 
the data is unable to distinguish between \sonic and \supersonic.

\begin{table}
	\caption{\label{tab:dndeta} 
The charged particle multiplicity ($dN_{ch}/d\eta$) at midrapidity for 
central \dau collisions at \sqsn~=~200, 62.4, 39, and 19.6 GeV.}
\begin{ruledtabular} \begin{tabular}{ccccc}
	\sqsn [GeV] & centrality & data & \ampt & (super)\sonic \\
	\hline
	200  &  0\%--5\% & 20.3$\pm$1.5 & 19.3 & 20.2$\pm$2 \\
	62.4 &  0\%--5\% & 12.4$\pm$2.4 & 16.1 & 15.0$\pm$2 \\
	39   & 0\%--10\% &  9.3$\pm$1.6 & 14.0 & 11.6$\pm$2 \\
	19.6 & 0\%--20\% &  5.8$\pm$1.1 &  9.7 & 9.7$\pm$2 \\
	\end{tabular} \end{ruledtabular}
\end{table}

Figure~\ref{fig:v2_theory}(e) includes hydrodynamic predictions of the 
$\eta$ dependence of $v_2$ in \dau collisions at \sqsn~=~200 GeV from 
Bozek and Broniowski~\cite{Bozek:2014cya}. These calculations utilize MC 
Glauber initial conditions, evolved with event-by-event $3+1$D viscous 
hydrodynamics, followed by statistical hadronization at freeze-out. The 
calculations are in good agreement with the data for $\eta>-2$ but start 
to under predict the data in the region $-3<\eta<-2$.

\subsection{Comparison of \dndeta results with \ampt calculations}
\label{sec:disc_dndeta}

\begin{figure*}[htbp]
\begin{minipage}{0.99\linewidth}
	\includegraphics[width=0.99\textwidth]{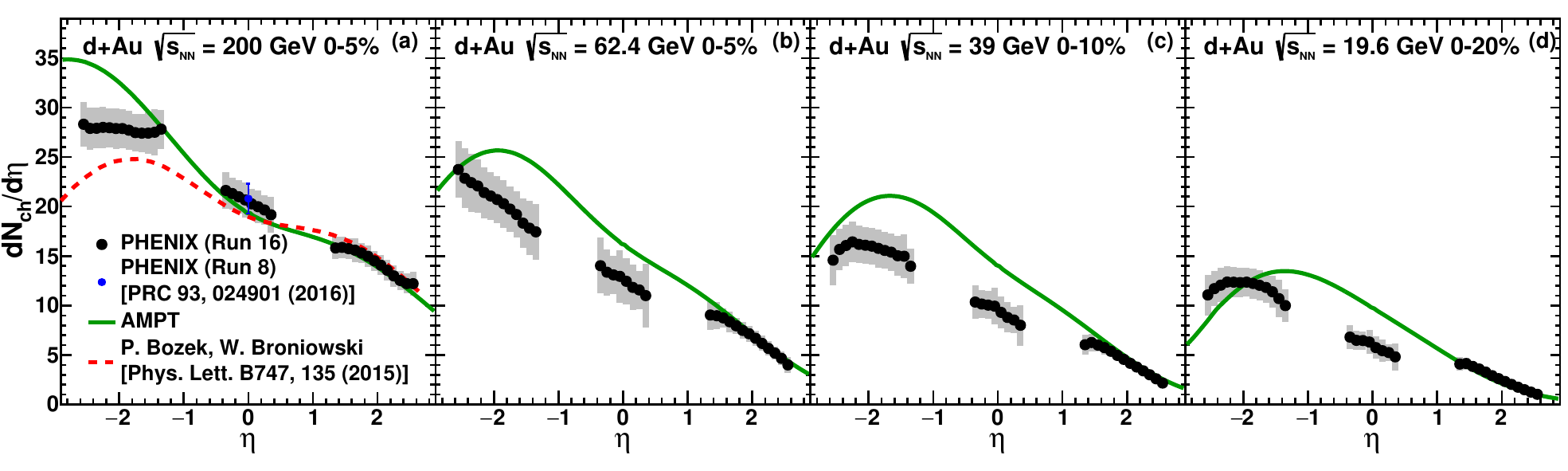}
	\caption{\label{fig:dndeta_energies} 
The \dndeta vs $\eta$ in central \dau collisions at \sqsn~=~(a) 200, 
(b) 62.4, (c) 39, and (d)~19.6~GeV. The solid [green] curves are the 
\ampt calculations in similar centrality bins.  The dashed [red] curve in 
panel (a) is a hydrodynamic prediction from Ref.~\cite{Bozek:2014cya} 
for 0\%--5\% central \dau collisions at \sqsn~=~200 GeV.}
\end{minipage}
\begin{minipage}{0.99\linewidth}
	\includegraphics[width=0.98\textwidth]{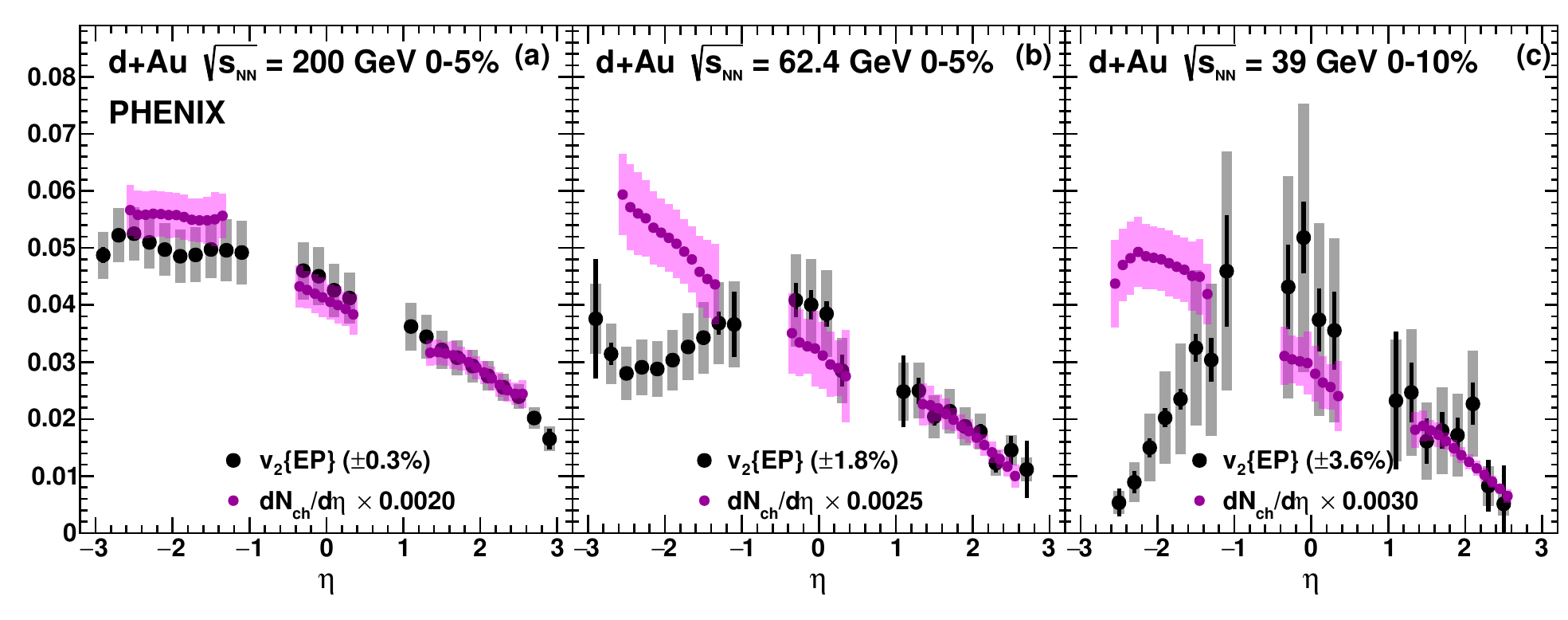}\\
	\includegraphics[width=0.98\textwidth]{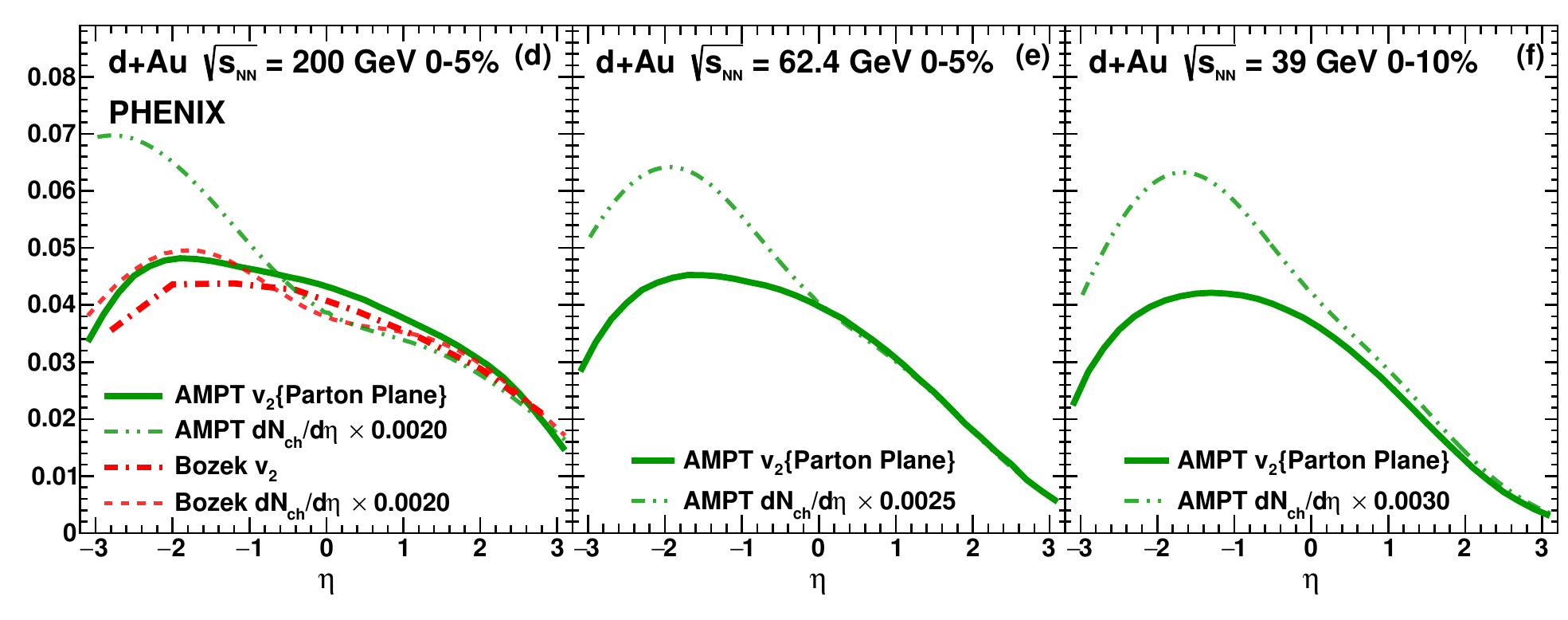}
	\caption{\label{fig:v2dndeta} 
The $v_2$ vs $\eta$ and \dndeta vs $\eta$, scaled to match the $v_2$ in 
$1.0<\eta<3.0$, for central \dau collisions at \sqsn~= (a) 200, 
(b) 62.4, and (c) 39 GeV. The dashed-double-dot and solid 
[green] curves in panels (d)--(f) show the results from \ampt using the 
same scaling factors determined from the data.  The dash and dash-dot 
[red] curves in panel (d) show the hydrodynamic predictions from 
Ref.~\protect\cite{Bozek:2014cya} in 0\%--5\% central \dau collisions at 
\sqsn~=~200 GeV, again using the same scaling factor determined from 
the data.}
\end{minipage}
\end{figure*}

The measurements of \dndeta vs $\eta$ in central \dau collisions at 
\sqsn~=~200, 62.4, 39, and 19.6 GeV are shown in 
Fig.~\ref{fig:dndeta_energies}. At all four energies, the \dndeta at 
backward rapidity is larger than that at forward rapidity, and the 
overall \dndeta decreases at all $\eta$ with decreasing energy. Also 
shown in Fig.~\ref{fig:dndeta_energies} are calculations from \ampt in 
the same centrality classes, as well as a prediction from Bozek and 
Broniowski~\cite{Bozek:2014cya} for 0\%--5\% central \dau collisions at 
\sqsn~=~200 GeV. At 200 GeV, \ampt agrees with the data well at mid and 
forward rapidities, while over-predicting the data at backward rapidity. 
The calculation from Bozek and Broniowski agrees with the data at mid to 
forward rapidity, while under-predicting the data at backward 
rapidities. It is worth noting that calculations from Bozek and 
Broniowski are substantially lower than the \ampt calculations for 
$\eta<-1$. This is potentially due to the centrality determination in 
\ampt (and data), which selects on multiplicity in the region 
$-3.9<\eta<-3.1$, which may naturally cause an autocorrelation with the 
\dndeta in the region $-3<\eta<-1$. At the lower three energies, \ampt 
matches the data well at forward psuedorapidity only and over-predicts 
the data at midrapidity.

We next turn to investigating whether there is a scaling of 
$v_2\propto\dndeta$. Figure~\ref{fig:v2dndeta}(a)--(c) shows the 
measured $v_2(\eta)$ overlaid with the \dndeta, where the \dndeta is 
arbitrarily scaled at each energy to match the $v_2$ at forward 
rapidity. We have chosen to match the \dndeta to the $v_2$ at $\eta>0$, 
as we expect the $v_2$ in this region to have the lowest contribution 
from nonflow, as discussed in Sec.~\ref{sec:ampt}. The required scaling 
factor increases with decreasing energy, with scaling factors of 0.0020, 
0.0025, and 0.0030 at 200, 62.4, and 39 GeV, respectively.

Figure~\ref{fig:v2dndeta}(d)--(f) shows the $\vtpp$ from \ampt overlaid 
with the scaled \dndeta, also from \ampt, using the same scaling factors 
determined from data. Additionally, Fig.~\ref{fig:v2dndeta}(d) shows the 
overlay of the calculations of $v_2$ and \dndeta from Bozek and 
Broniowski, where \dndeta is scaled by the same factor of 0.0020.

Starting with the 200 GeV results in Fig.~\ref{fig:v2dndeta}(a)\&(d), we 
find that when using a constant scaling factor across $\eta$, the scaled 
\dndeta and $v_2(\eta)$ agree well within uncertainties. The increase in 
the $v_2$ from forward to backward rapidity is matched by the increase 
in the \dndeta. In comparison, the \ampt shows an approximate scaling 
only at forward rapidity, although a better match is found when using a 
scaling factor of 0.0022, rather than 0.0020. The scaled \dndeta breaks 
from the $\vtpp$ for $\eta<1$, indicating that within \ampt there is no 
direct scaling of the \dndeta and $\vtpp$. Similarly, the calculations by
Bozek and Broniowski show an approximate scaling at forward rapidity, 
and a modest scale breaking at backward rapidities.

At 62.4 and 39 GeV, we find that the scaled \dndeta and $v_2(\eta)$ 
agree within uncertainties at mid and forward rapidities. At backward 
rapidity however, the scaled \dndeta is significantly larger than the 
$v_2$ for the same scaling factor. It is notable that \ampt $v_{2}$ 
does not scale with \dndeta at backward rapidity at any energy. As 
discussed in Sec.~\ref{sec:ampt}, \ampt calculations indicate that 
there could be an anti-correlation effect at backward rapidity that 
decreases the observed $v_2$ relative to the true $v_2$ when using the 
event-plane method. Further investigations into potential nonflow 
anti-correlations in the event-plane method with a small $\Delta\eta$ 
gap would be useful to shed more light on these possible conclusions.

\section{Summary and conclusions}
\label{sec:conclusions}

PHENIX has presented new measurements of the second order flow 
coefficient $v_2$ in bins of centrality in \dau collisions at 
\sqsn~=~200, 62.4, 39, and 19.6 GeV as a function of \pt and $\eta$. We 
find that at mid to forward rapidities and low \pt, $v_2$ appears to be 
dominated by flow, where we define flow as the translation of initial 
geometry to final-state momentum anisotropy via interactions between 
medium constituents. In contrast, at backward rapidity and high \pt, 
nonflow becomes an increasingly significant contribution.

It would be interesting to compare the $v_2$ results measured in the 
\dau beam energy scan with those measured in \pp and \ppb collisions at 
the LHC. The multiplicity ranges probed in the \dau beam energy scan are 
comparable to those in \pp collisions at the LHC, which range from 
$dN_{ch}/d\eta\approx4$ in MB collisions to $dN_{ch}/d\eta>80$ 
in very high-multiplicity events~\cite{Adam:2015gka}. 
Comparing the different systems at similar multiplicities, but vastly 
different collision energies and initial geometries, may give further 
insight into the underlying mechanism generating the $v_2$ signal.
We further present measurements of \dndeta vs $\eta$ at all four 
energies. At 200 GeV, we find that a constant scale factor yields 
agreement between the measured $v_2$ vs $\eta$ and the shape of \dndeta. 
At 62.4 and 39 GeV, the shapes of $v_2$ and \dndeta match well at mid 
and forward rapidity, however the \dndeta increases at backward rapidity 
while the $v_2$ decreases. This presents a different picture than that 
observed at 200 GeV, and may be due to anti-correlations present in the 
event-plane method when the $\Delta\eta$ gap becomes small.

These results provide further evidence that the $v_2$ measured in small 
systems arises from initial geometry coupled to interactions between 
medium constituents, whether described by parton scattering or 
hydrodynamics. In \dau collisions at \sqsn~=~200 GeV, these flow effects 
dominate and they continue to play a significant, though less dominant 
role all the way down to \sqsn~=~19.6 GeV.


\section*{ACKNOWLEDGMENTS}  

We thank the staff of the Collider-Accelerator and Physics
Departments at Brookhaven National Laboratory and the staff of
the other PHENIX participating institutions for their vital
contributions.  We acknowledge support from the 
Office of Nuclear Physics in the
Office of Science of the Department of Energy,
the National Science Foundation, 
Abilene Christian University Research Council, 
Research Foundation of SUNY, and
Dean of the College of Arts and Sciences, Vanderbilt University 
(U.S.A),
Ministry of Education, Culture, Sports, Science, and Technology
and the Japan Society for the Promotion of Science (Japan),
Conselho Nacional de Desenvolvimento Cient\'{\i}fico e
Tecnol{\'o}gico and Funda\c c{\~a}o de Amparo {\`a} Pesquisa do
Estado de S{\~a}o Paulo (Brazil),
Natural Science Foundation of China (People's Republic of China),
Croatian Science Foundation and
Ministry of Science and Education (Croatia),
Ministry of Education, Youth and Sports (Czech Republic),
Centre National de la Recherche Scientifique, Commissariat
{\`a} l'{\'E}nergie Atomique, and Institut National de Physique
Nucl{\'e}aire et de Physique des Particules (France),
Bundesministerium f\"ur Bildung und Forschung, Deutscher
Akademischer Austausch Dienst, and Alexander von Humboldt Stiftung (Germany),
J. Bolyai Research Scholarship, EFOP, the New National Excellence
Program ({\'U}NKP), NKFIH, and OTKA (Hungary),
Department of Atomic Energy and Department of Science and Technology (India), 
Israel Science Foundation (Israel), 
Basic Science Research Program through NRF of the Ministry of Education (Korea),
Physics Department, Lahore University of Management Sciences (Pakistan),
Ministry of Education and Science, Russian Academy of Sciences,
Federal Agency of Atomic Energy (Russia),
VR and Wallenberg Foundation (Sweden), 
the U.S. Civilian Research and Development Foundation for the
Independent States of the Former Soviet Union, 
the Hungarian American Enterprise Scholarship Fund,
the US-Hungarian Fulbright Foundation,
and the US-Israel Binational Science Foundation.


\section*{Appendix A:  Centrality dependence of $v_2(\pt)$}
\label{appendix:A}

\begin{figure*}[htbp]
\begin{minipage}{0.98\linewidth}
    \includegraphics[width=0.8\linewidth]{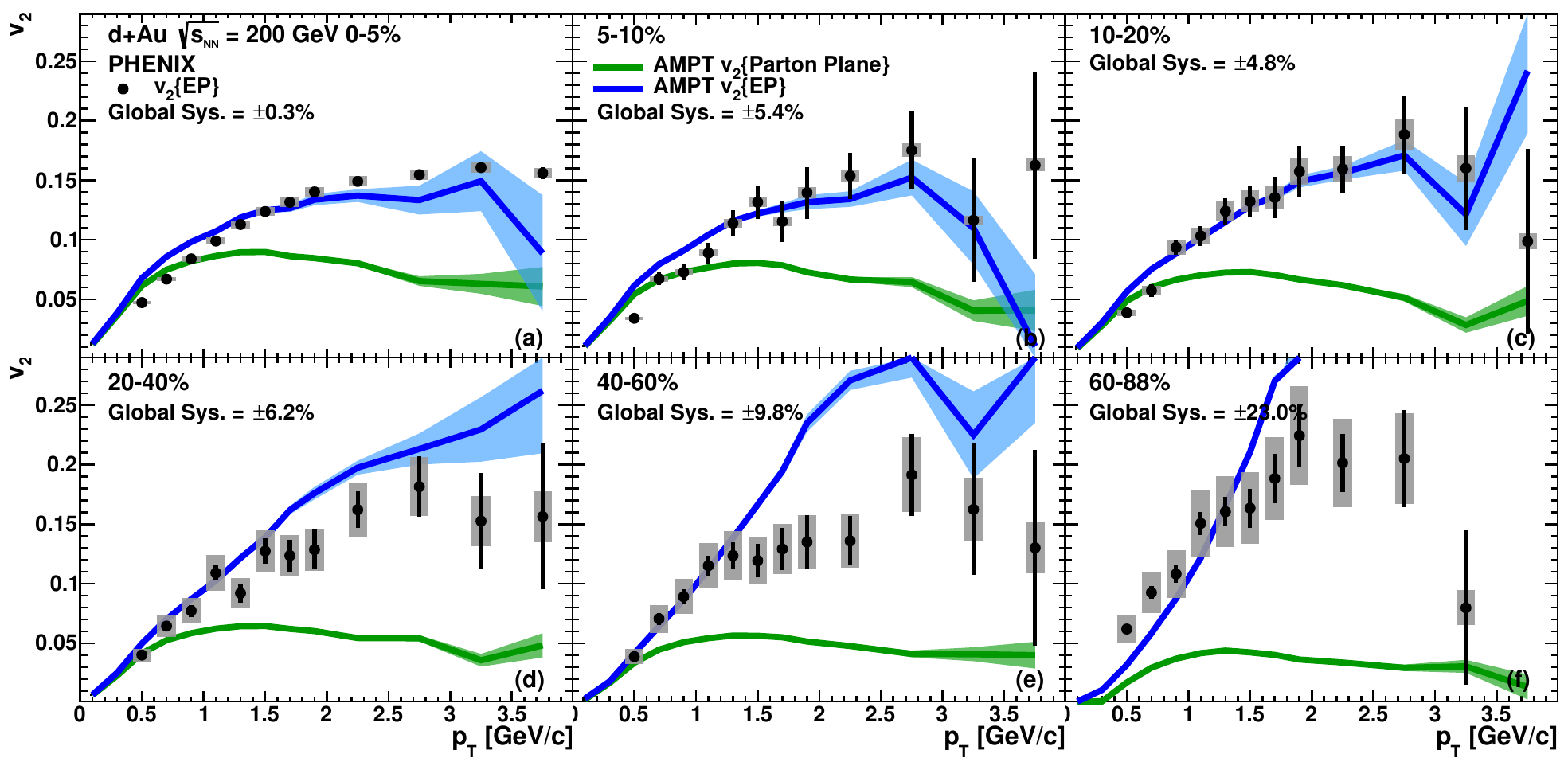}
\caption{\label{fig:v2pt_cent200} 
For \dau collisions at \sqsn~=~200 GeV, the value of $v_2$ as a function 
of \pt in (a) 0\%--5\%, (b) 5\%--10\%, (c) 10\%--20\%, (d) 20\%--40\%, 
(e) 40\%--60\%, and (f) 60\%--88\%.  The upper [blue] curves show 
calculations from the \ampt model~\cite{Lin:2004en}, where the $v_2$ is 
calculated using the event-plane method, as described in the text.  The 
lower [green] curves show \ampt calculations, where the $v_2$ is 
calculated relative to the parton plane.
}
\end{minipage}
\begin{minipage}{0.98\linewidth}
    \includegraphics[width=0.8\linewidth]{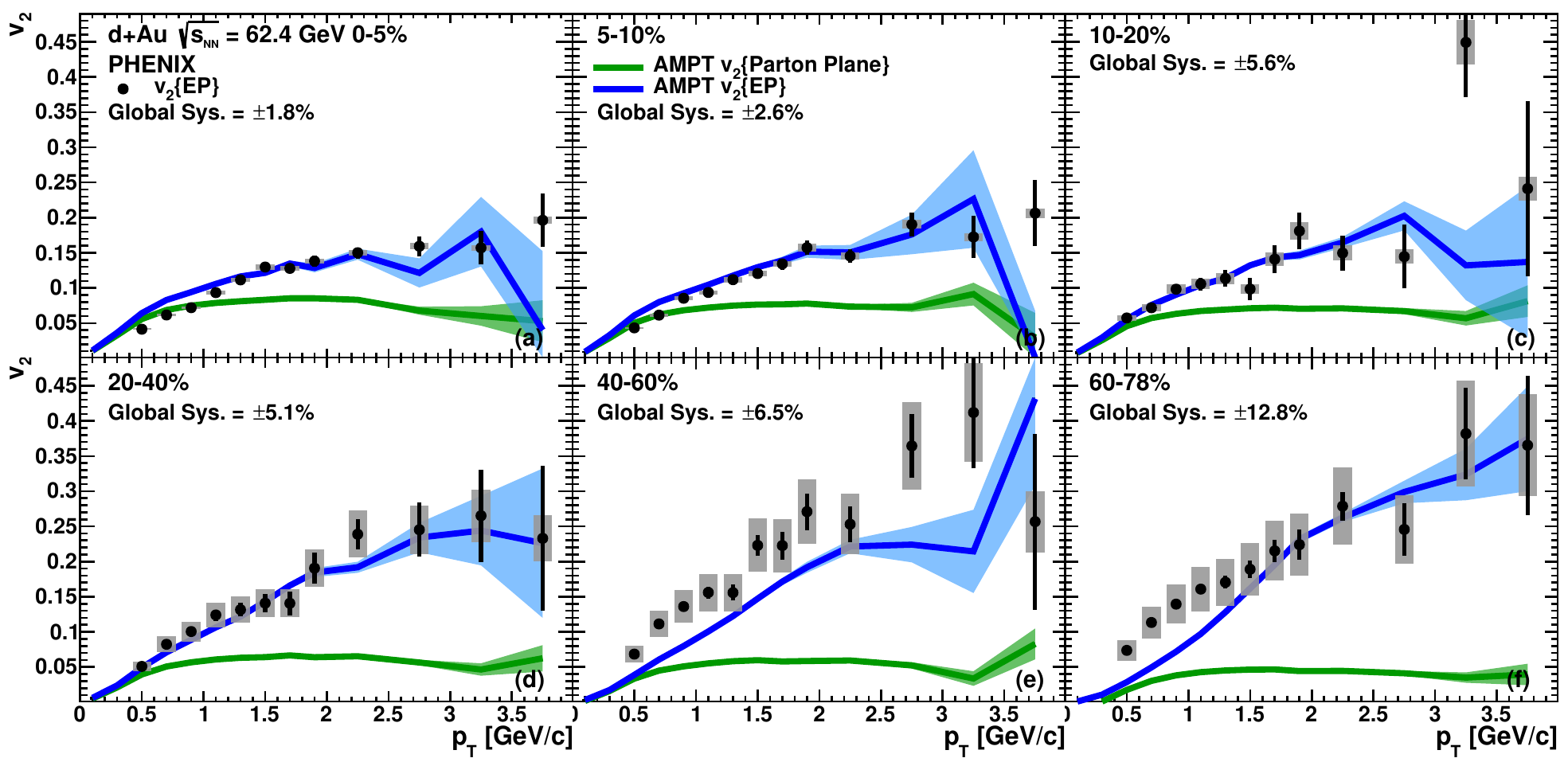}
\caption{\label{fig:v2pt_cent62} 
For \dau collisions at \sqsn~=~62.4 GeV, descriptions of the 
symbols and curves are the same as in Fig.~\ref{fig:v2pt_cent200}.} 
\end{minipage}
\begin{minipage}{0.98\linewidth}
    \includegraphics[width=0.8\linewidth]{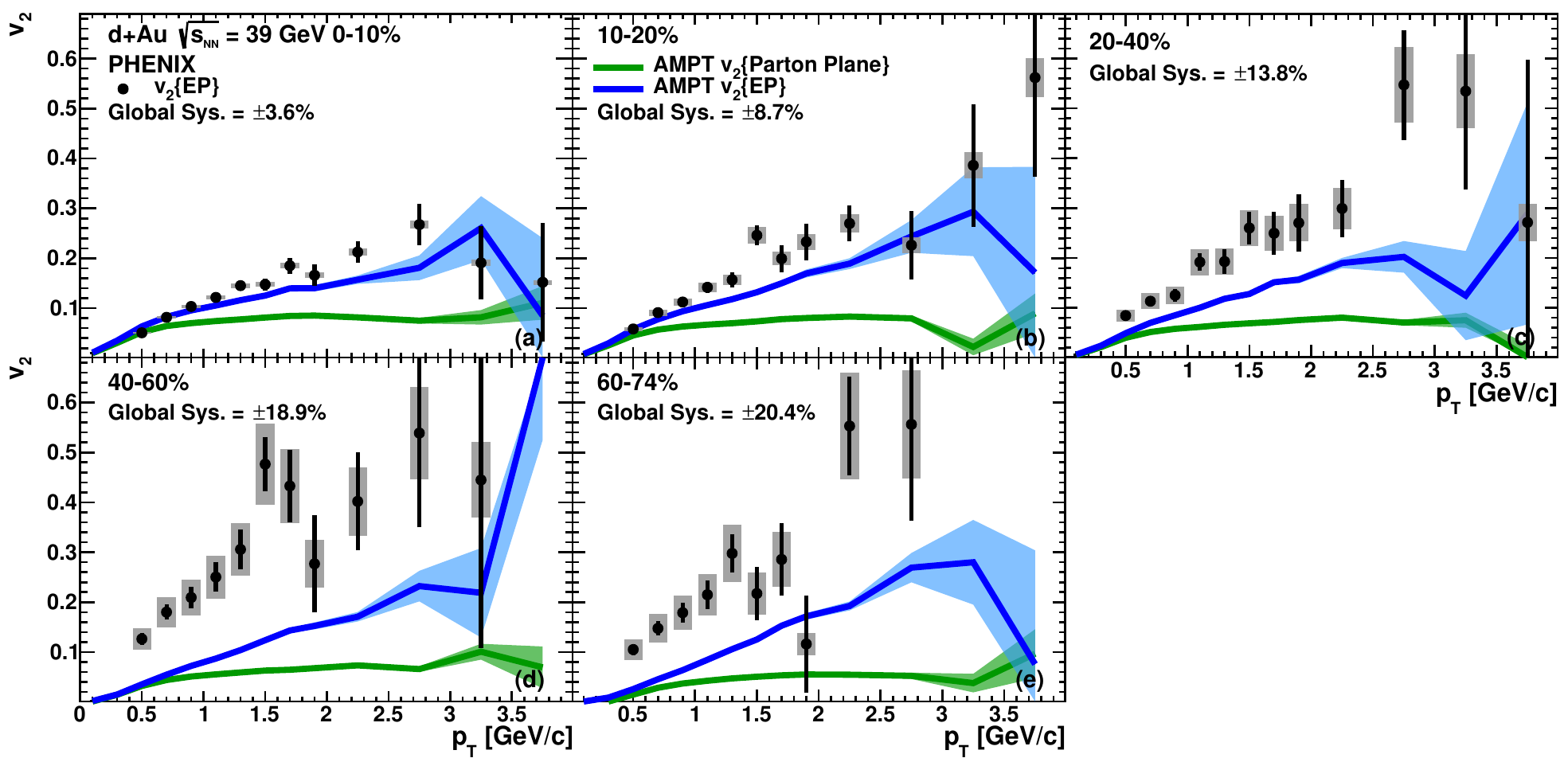}
\caption{\label{fig:v2pt_cent39} 
For \dau collisions at \sqsn~=~39 GeV, descriptions of the 
symbols and curves are the same as in Fig.~\ref{fig:v2pt_cent200}.} 
\end{minipage}
\end{figure*}

The $v_2(\pt)$ in centrality bins for \dau collisions at \sqsn~=~200, 
62.4, and 39 GeV are shown in Figs.~\ref{fig:v2pt_cent200}, 
\ref{fig:v2pt_cent62}, and \ref{fig:v2pt_cent39}, respectively.

\clearpage

\section*{Appendix B: \ampt details}
\label{appendix:B}

The \ampt calculations shown in this work are generated following 
Ref.~\cite{Koop:2015trj}. We use \ampt Version 2.26, which is 
additionally modified to utilize the Hulth\'en wavefunction description 
of the deuteron and black disk nucleon-nucleon interactions with the 
Monte-Carlo Glauber component. The input \ampt parameters which are 
tuned outside the default values are shown in Table~\ref{tab:ampt}. 
Unlike Ref.~\cite{Koop:2015trj}, which uses a parton interaction cross 
section of 1.50 mb, we use a parton interaction cross section of 
$\sigparton=0.75$ mb, as we find it provides a better description of the 
centrality binned data.

\begin{table}
\caption{\label{tab:ampt} Nondefault parameter values used when running 
\ampt.}
\begin{ruledtabular}	\begin{tabular}{ll}
	Parameter & Value \\
	\hline
	ISOFT & 4 \\
	PARJ(41) & 2.2 \\ 
	PARJ(42) & 0.5 \\ 
	Parton screening mass & $6.4528d0$ (0.75 mb) \\
	alpha in parton cascade & $0.47140452d0$ \\
	ihjsed & 11 \\	
	\end{tabular} \end{ruledtabular}
\end{table}

In addition to the full \ampt calculations with both partonic and 
hadronic scattering shown in Figs.~\ref{fig:v2_ampt} 
and~\ref{fig:v2pt_cent200}--\ref{fig:v2pt_cent39}, we provide 
calculations for the following three cases:
\begin{itemize}
	\item \textbf{N.S.} -- Both partonic scattering and hadronic 
scattering turned off (i.e. no scattering)
	\item \textbf{P.S.} -- Partonic scattering only
	\item \textbf{H.S.} -- Hadronic scattering only
\end{itemize}

\begin{figure*}[htbp]
\begin{minipage}{0.99\linewidth}
        \includegraphics[width=0.99\linewidth]{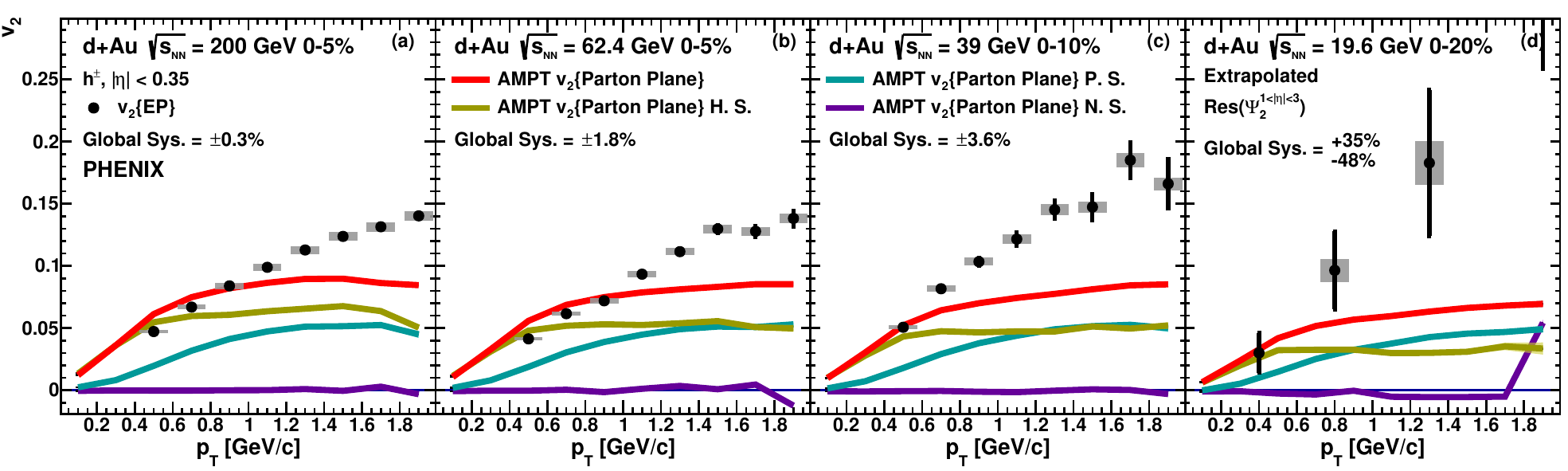}\\
        \includegraphics[width=0.99\linewidth]{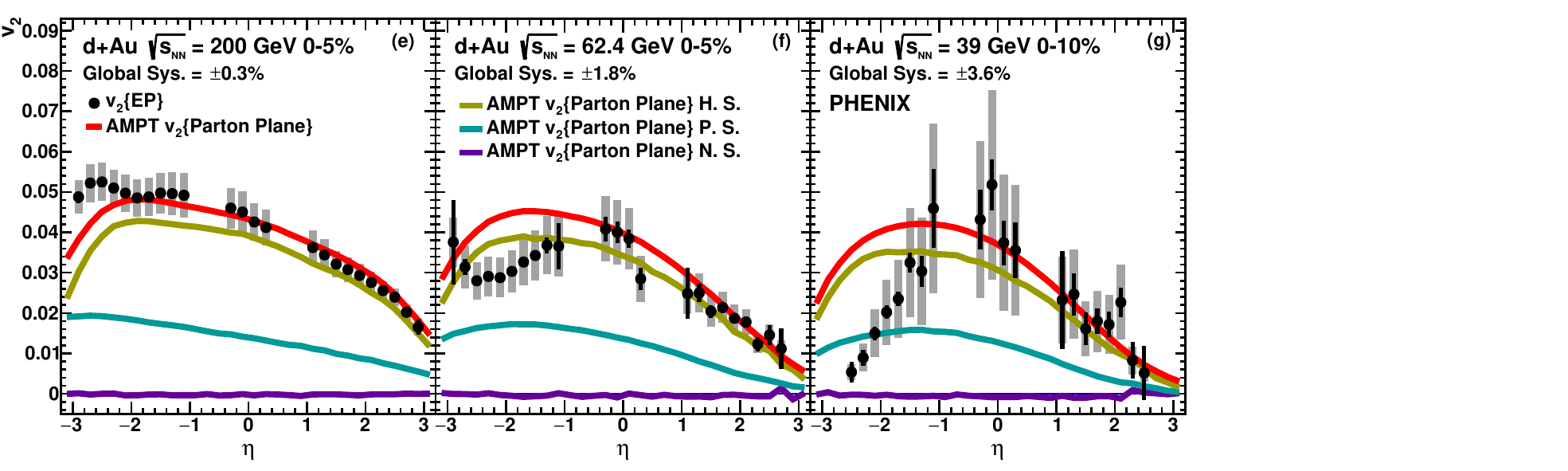}
	\caption{\label{fig:v2_ampt_scat_pp}
(a, b, c, d) the value of $v_2$ vs $p_T$ in central~\dau collisions 
at  \sqsn~=~200, 62.4, 39, and 19.6 GeV.  
(e, f, g) the value of  $v_2$ vs   $\eta$ 
in central  \dau collisions at  \sqsn~=~200, 62.4, and 39~GeV. 
The curves are calculations from  \ampt under different conditions. 
With ordering of curves from top to bottom (a, b, c, d) at  $p_T=0.6$
and (e, f, g) at  $\eta=0$, the uppermost [red] curve is  \ampt with both
partonic and hadronic scattering; the upper-middle [yellow] curve 
is  \ampt with hadronic scattering only (H.S.); the lower-middle [cyan] curve 
is  \ampt with partonic scattering only (P.S.); and the lowest [purple] curve
is  \ampt with no scattering (N.S.). For all  \ampt curves, the  $v_2$ is
calculated relative to the initial parton plane.
}
\end{minipage}
\begin{minipage}{0.99\linewidth}
        \includegraphics[width=0.99\linewidth]{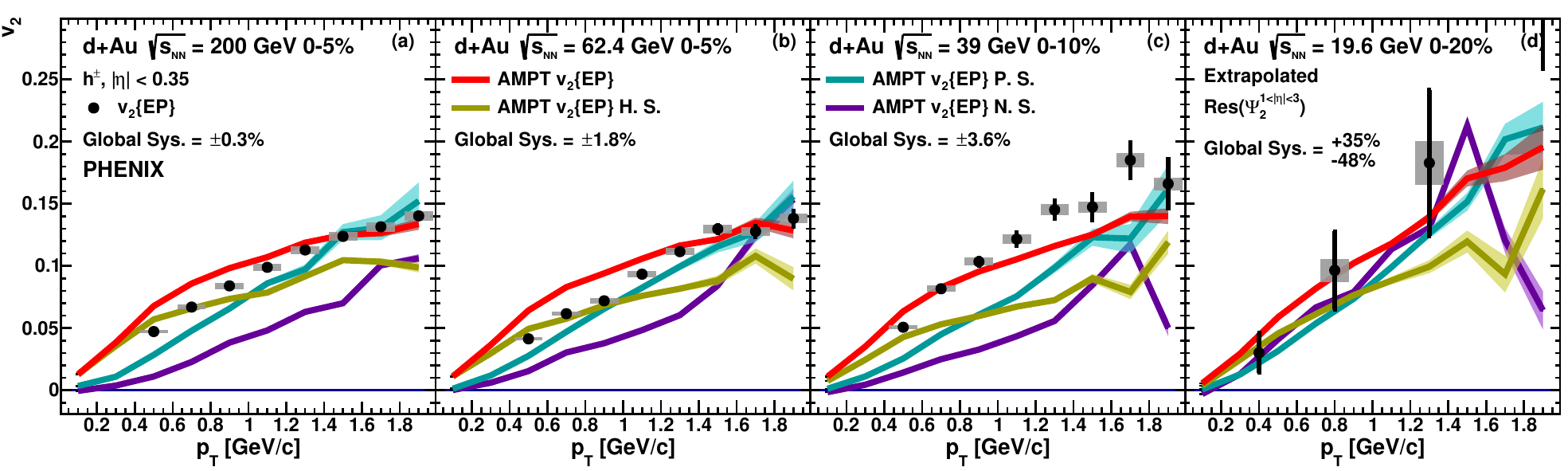}\\
        \includegraphics[width=0.99\linewidth]{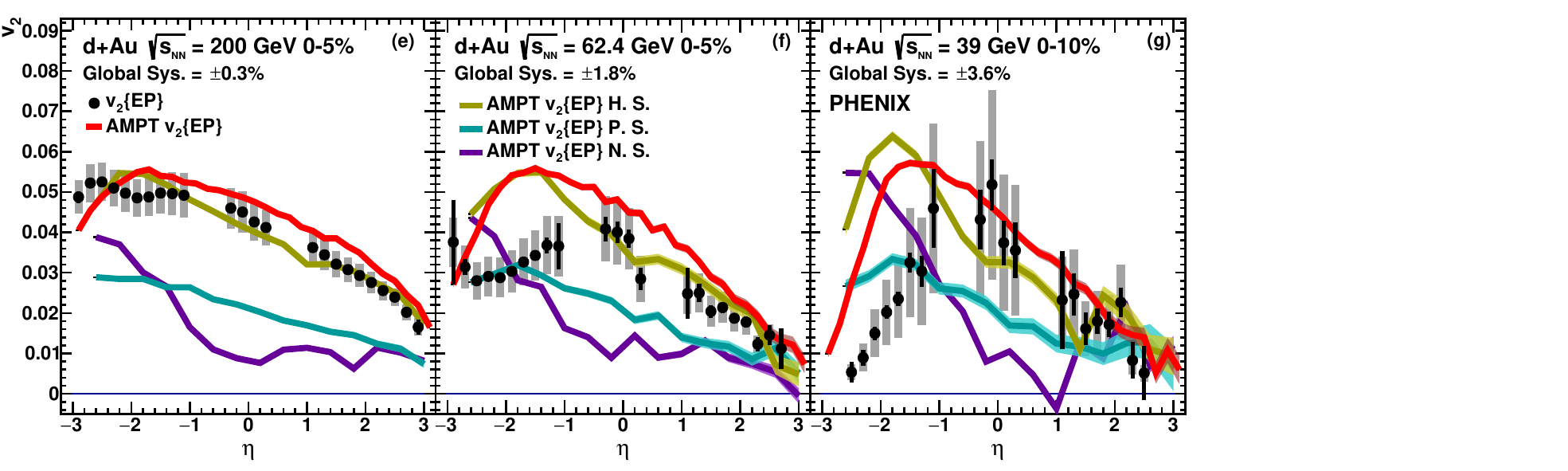}
	\caption{\label{fig:v2_ampt_scat_ep}
The description of all symbols and curves are the same as in
Fig.~\ref{fig:v2_ampt_scat_pp}, except that for all \ampt curves
the $v_2$ is calculated relative to the final-state event plane. 
}
\end{minipage}
\end{figure*}

To turn off hadronic scattering we turn off the hadron cascade 
(NTMAX~=~3). In order to turn off partonic scattering we set the parton 
interaction cross section to 0 mb. Figures~\ref{fig:v2_ampt_scat_pp} 
and~\ref{fig:v2_ampt_scat_ep} show the results for central \dau 
collisions.

Figure~\ref{fig:v2_ampt_scat_pp} shows the results from \ampt for $v_2$ 
as a function of \pt and pseudorapidity using the parton plane method, 
which yields a pure flow result with respect to initial geometry.  
Focusing on the \pt dependence in Fig.~\ref{fig:v2_ampt_scat_pp} (upper 
panels), the hadronic scattering only scenario results in larger $v_2$ 
compared to the partonic scattering only scenario at low $\pt<1$ GeV/$c$ 
and then a comparable $v_2$ for higher \pt.  Note that these 
contributions cannot simply be summed to achieve the result with both 
partonic and hadronic scattering because the space-time input for the 
hadronic scattering stage changes depending on whether there is or is no 
partonic scattering stage.  The significantly larger $v_2$ in the 
hadronic scattering only scenario at low-\pt is most clearly seen in 
Fig.~\ref{fig:v2_ampt_scat_pp} (lower panels) because the $v_2$ as a 
function of pseudorapidity is integrated over all \pt.

At high-\pt, the partonic-scattering-only scenario has a more comparable 
contribution to the hadronic-scattering-only scenario, with it being 
slightly smaller at 200 GeV and slightly larger at 39 GeV.  Because the 
\ampt model employs a formation time for partons such that higher \pt 
partons start scattering earlier in time, it makes sense that this 
contributes more significantly.  It is notable that in 
Ref.~\cite{Koop:2015trj}, it was shown that the parton scattering began 
to dominate for $\pt>0.8$ GeV/$c$. This difference is likely due to the 
larger parton interaction cross section of 1.50 mb used in 
Ref.~\cite{Koop:2015trj}. As the collision energy decreases, the 
partonic scattering contributes more to the overall $v_2$ signal. As 
discussed in Sec.~\ref{sec:ampt}, the no scattering case has $\vtpp=0$ 
by definition, as it no longer has the ability to translate initial 
geometry to momentum anisotropy.

Figure~\ref{fig:v2_ampt_scat_ep} shows results calculated using 
the event-plane method ($\vtep$), i.e. simulating the experimental 
method of extracting $v_2$. The general statement above that hadronic 
scattering dominates at low-\pt while partonic scattering contributes 
mainly at higher \pt remains true down to \sqsn~=~39 GeV. However, as 
discussed in Sec.~\ref{sec:ampt}, the case with both partonic and 
hadronic scattering turned off now shows a nonzero $\vtep$ signal. This 
$\vtep$ result without scattering indicates that nonflow is small at 
low-\pt but grows with increasing \pt. For collision energies of 39 GeV 
and above, the $\vtep$ result without scattering is inconsistent with 
the measured results as a function of both \pt and $\eta$. However, at 
\sqsn~=~19.6 GeV, the $\vtep$ results in all four cases are nearly 
consistent. This appears to indicate that, within \ampt, the $\vtep$ 
measurement is dominated by nonflow contributions and does not reflect 
the true flow even at low \pt.

\clearpage



%
 
\end{document}